\documentstyle[12pt,epsf]{article}

\newcommand{\df}{istribution function}
\newcommand{\cf}{oefficient function}
\newcommand{\een}{\end{equation}}
\newcommand{\bee}{\begin{equation}}
\newcommand{\ee}{\end{equation}}
\newcommand{\be}{\begin{equation}}

\newcommand{\alfs}{\alpha_s}
\newcommand{\beqn}{\begin{eqnarray}}
\newcommand{\eeqn}{\end{eqnarray}}

 \newcommand{\rcite}[1]{{\cite{#1}}}
 \newcommand{\rref}[1]{{(\ref{#1})}}
 
 \newcommand{\rlabel}[1]{{\label{#1}}}
 \newcommand{\rbibitem}[1]{\bibitem{#1}}

\def\cjp#1, #2, #3 {C.~J.~Phys.\underline{#1}~(19#3)~ #2}
\def\ajp#1, #2, #3 {Am.~J.~Phys.\underline{#1}~(19#3)~ #2}
\def\pr#1, #2, #3 {Phys.~Rev.~\underline{D#1}~(19#3)~ #2}
\def\pl#1, #2, #3 {Phys.~Lett.~\underline{#1B}~(19#3)~ #2}
\def\mpla#1, #2, #3{Mod.~Phys.~Lett.~\underline{A#1} #2~(19#3)}
\def\np#1, #2, #3  {Nucl.~Phys.~\underline{B#1}~(19#3)~ #2}
\def\sjnp#1, #2, #3  {Sov.~Jour.~Nucl.~Phys.~\underline{#1}~(19#3)~#2}
\def\npps#1, #2, #3{Nucl.~Phys.~\underline{B#1} (Proc. Suppl),
                     #2---#3~(19#3)}
\def\zp#1, #2, #3 {Z.~Phys.~\underline{C#1}~(19#3)~ #2}

\begin{document}

\begin{titlepage}
\begin{flushright}
hep-ph/9707367 \\
NORDITA 97/58 \\
\today \\
July 1997 \\
revised version \\
\end{flushright}

\vfill

\begin{center}
\begin{bf}
{\Large \bf
     Resummation of  $(-b_0\alfs)^n$ corrections to the photon-meson
     transition form factor $\gamma^{\ast}+ \gamma
     \rightarrow \pi^0 $} \\[2cm] \end{bf}

P. Gosdzinsky$^a$\footnote{ email: gosdzins@nordita.dk}
 and
N. Kivel$^b$\footnote{ email: kivel@thd.pnpi.spb.ru} \\[1cm]
$^a$NORDITA, Blegdamsvej 17, DK-2100 Copenhagen \O , Denmark \\
$^b$Petersburg Nuclear Physics Institute, 188350, Gatchina, Russia \\

\vfill

\end{center}
\vfill
\begin{abstract}

We have resummed all the
$(-b_0\alfs)^n$ contributions to the photon-meson
transition form factor $F_{ \pi \gamma  }$. To do this, we
have used the assumption of `naive nonabelianization' (NNA).
Within NNA, a series in $(N_f\alfs)^n$ is interpreted as a
series in $(-b_0\alfs)^n$ by means of the restoration of the
full first QCD $\beta$-function coefficient $-b_0$ by hand.  We
have taken into account corrections to the leading order c\cf {}
and to the evolution of the d\df{}. Due to conformal symmetry
constraints, it is possible to find the eigenfunctions of the
evolution kernel. It turns out that the nondiagonal corrections
are small, and neglecting them we obtained a representation for
the d\df {} with multiplicatively renormalized moments.  For a
simple shape of the d\df{}, which is close to the asymptotic
shape, we find that the radiative correction decrease the LO by
$30\%$, and the uncertainty in the resummation lies between
$10\%$ and $2\%$ for $Q^2$ between 2 and 10 GeV$^2$.

\end{abstract}
\vfill
\end{titlepage}

\section{Introduction}

We can expect that perturbative QCD (pQCD) works well to describe the
process \\ $ \gamma ^* (q1) \gamma (q2) \rightarrow \pi ^0$ for
accessible values of $q_1^2 <0$ and $q_2^2 \leq 0$. The form
factor for this process is given by \rcite{BL}
\be F_{\pi \gamma } = \int _0 ^1 \varphi_\pi (x,\mu^2) {\cal C}
(x,q_1,q_2,\mu^2) dx + \ldots \rlabel{defF}
\ee
The leading
term in a $1/Q^2$ expansion, being $Q^2$ the momentum transfer,
is given by the integral in \rref{defF}. The dots, $\ldots$, stand for
higher twist contributions, which are subleading in the $1/Q^2$
expansion.

The coefficient function $ {\cal C} (x,q_1,q_2) $, which accounts for
the transition from photons to quarks, can be completely described
within pQCD. It is known to one loop, and a detailed
analysis can be found for example in \rcite{braaten,KMR,PACO}.

The distribution function $\varphi (x, \mu ^2)$ can be
interpreted as the
transition probability of a $\pi ^0 $ with momentum $P$
into two quarks with momenta
$xP$ and $(1-x)P$ respectively. Only its evolution with $\mu ^2$ is
given by pQCD, but not its shape, or $x$ dependence. Here, the situation
is rather unclear, and there exist contradictory statements in the
literature. Due to our present inability to extract
it directly from experiment, we can only make
some choice, (or guess) for $\varphi (x, \mu ^2)$. The two most
popular choices are the ``asymptotic wave function'',
$\varphi _{as} = 6 x (1 -x )$, and the CZ model,
$\varphi _{CZ} = 30 x (1-x) (1-2x)^2$. Very recently, new
experimental data have appeared, and more are expected, \rcite{CLEO}.
It is expected that this will allow to constrain,
and perhaps even extract
with some accuracy, the distribution function $\varphi (x, \mu ^2)$.
In fact, one of the major goals of the study of the form factor
$F_{\pi \gamma  }$ is precisely to obtain more information
on the distribution function.

In order to extract as much information as possible from the
experimental data, precise theoretical predictions for the
coefficient function, beyond the one loop level, are needed,
and it might even be necessary to include nonleading contributions
in the $1/Q^2$ counting. In this work, we analyze the
accuracy within which the leading order in $1/Q^2$, that is, the
integral in \rref{defF}, can predict the form factor
$F_{\pi \gamma   }$ within Naive Nonabelianization, NNA.

The idea of NNA is based on the observation that corrections related
to the evolution with the coupling can represent a source of
potentially large perturbative coefficients. The extraction
of these large contributions can give important information on
higher order corrections. In QCD, this extraction can be done by
evaluating the relevant feynman diagrams to leading order in the large
$N_f$ limit, and interpreting the series in $(N_f \alpha _s)$ as a
series in $(-b_0 \alpha _s)$, restoring the full QCD $\beta$-function
coefficient $-b_0$ by hand
\footnote{Here, for the $\beta-$ function,
we adopt
$\beta=-b_0\alfs^2+\cdots, \
b_0=\frac{1}{4\pi}(\frac{11}{3}N_c-\frac{2}{3}N_f$)}.
Techniques to perform a summation of these large coefficients
have been developed in \rcite{BB2,BBB}.

In this work, we will only calculate leading twist corrections, and use
the ultraviolet dominance assumption \rcite{NNAbraun} to estimate
the higher twist corrections. This assumption is based on the
observation that due to the (infrared) renormalon ambiguity, the leading
twist result is affected by an intrinsic ambiguity,

\be
\delta C_{LT} \ \ \ \alpha \ \ \ A(y)
\left( {\Lambda \over Q } \right) ^2 \,
\rlabel{intAMB}
\ee
where $A(y)$ is a calculable function
that is completely fixed by the residues of
the Borel integrand, see below, and $y$ denotes the variables
on which it depends. The ambiguity in \rref{intAMB} is cancelled by
another (ultraviolet) renormalon ambiguity in the higher twist
contributions. According to the ultraviolet dominance assumption,
not only the ambiguity, but the whole higher twist contribution is
proportional to $A(y)$.

This work is organized as follows: In the next section, we briefly
review the current state of art. In section 3, we compute the
coefficient function within NNA. We find that the c\cf {} has
two $IR$-renormalon poles and that when one of the
photons is on
shell, there are additional ambiguities coming from the region
$x \rightarrow 0$ and $x \rightarrow 1$. These new ambiguities, which
are related to the infrared
region, lead to new power corrections to the form factor.
In section 4, the NNA evolution
kernel is presented. We obtain an expansion for the
d\df {}  in series of Gegenbauer
polynomials with the upper index shifted by $b_0\alfs$. We find that the
nondiagonal part of the anomalous dimension matrix in this basis is
much smaller than the diagonal part. In a first approximation,
 we can neglect
%%%%%%%%%%%%%%%%%%%%%%%%%%%%%%%%%%%%%%%%%%%%%%%%%%
% CHANGE BEGINS
%%%%%%%%%%%%%%%%%%%%%%%%%%%%%%%%%%%%%%%%%%%%%%%%%%
these nondiagonal terms, obtaining multiplicatively renormalized moments.
 In section 4, we also present two models for the wave functions, and
make some comments on the implications that
%%%%%%%%%%%%%%%%%%%%%%%%%%%%%%%%%%%%%%%%%%%%%%%%%%
% CHANGE EDNS
%%%%%%%%%%%%%%%%%%%%%%%%%%%%%%%%%%%%%%%%%%%%%%%%%%
conformal symmetry has.
In section 5 we obtain the final result for the form factor in the NNA
approximation. We present some numerical results for a
simplified d\df {}, where only the first term of the expansion is kept.
In this case, the
shape of the d\df\ is close to the asymptotic one.
In section 6 we present our conclusions, and finally,
 we present two appendices with technical details of the
calculations.

\section{The meson--photon transition form factor}

The transition from two photons to a $\pi^0$ meson,
\bee
\gamma^{\ast}(q_1)+\gamma (q_2) \rightarrow \pi^0(P)
\een
is described by the amplitude $T$
\bee
      T=e^{\alpha}(q_1)e^{\beta}(q_2)4\epsilon_{\alpha \beta \lambda \rho}
          P^{\lambda}\frac{1}{2}(q_1-q_2)^{\rho}
          F_{\pi\gamma}(Q^2,\omega).
\een
Here, $F_{\pi\gamma}(Q^2,\omega)$ is the photon--meson form factor,
$e^{\alpha}(q_1)\ \mbox{and}\ e^{\beta}(q_2)$ are the polarizations
of the colliding photons, \ $\ P=q_1+q_2,\
 -q_1^2>0,\ -q_2^2\geq 0,\ \ Q^2=-(q_1-q_2)^2/4 \ \mbox{and} \
\omega=(q_1^2-q_2^2)/(q_1^2+q_2^2) \leq 1$ is the parameter of asymmetry
 of the photons.
If one of the photons ($\gamma _2$) is real, $q_2^2\ =\ 0$ and
$\omega\ =\ 1$. In experimentally
accessible regions $\omega$ is very close to one.

In fig.\ref{regimes} we have represented
the process diagrammatically.  The dominant contribution is given by
fig\ref{regimes}.1, where a large virtual momentum flows through the
subgraph containing the two photon vertices.
The other regimes correspond to a long distance propagation in the
$q_2$ channel \rcite{MURA}.
The second regime
corresponds to the case where the large momentum flows through the
central block containing a large virtual photon fig.\ref{regimes}.2.
The third regime represents the situation when one of the quarks absorbs
a large virtual momentum and
carries almost all the momentum of the hadron
 and the second quark is soft, fig.\ref{regimes}.3.
Power counting predicts that the leading order for these contributions
is $1/Q^4$.
While for $\omega \sim 1$ all three regimes are important,
only the first one is relevant for the situation in which
both photons are off shell, $\omega < 1$. We will now discuss
this contribution.

\begin{figure}
\begin{center}
\leavevmode\epsfxsize=12cm\epsfbox{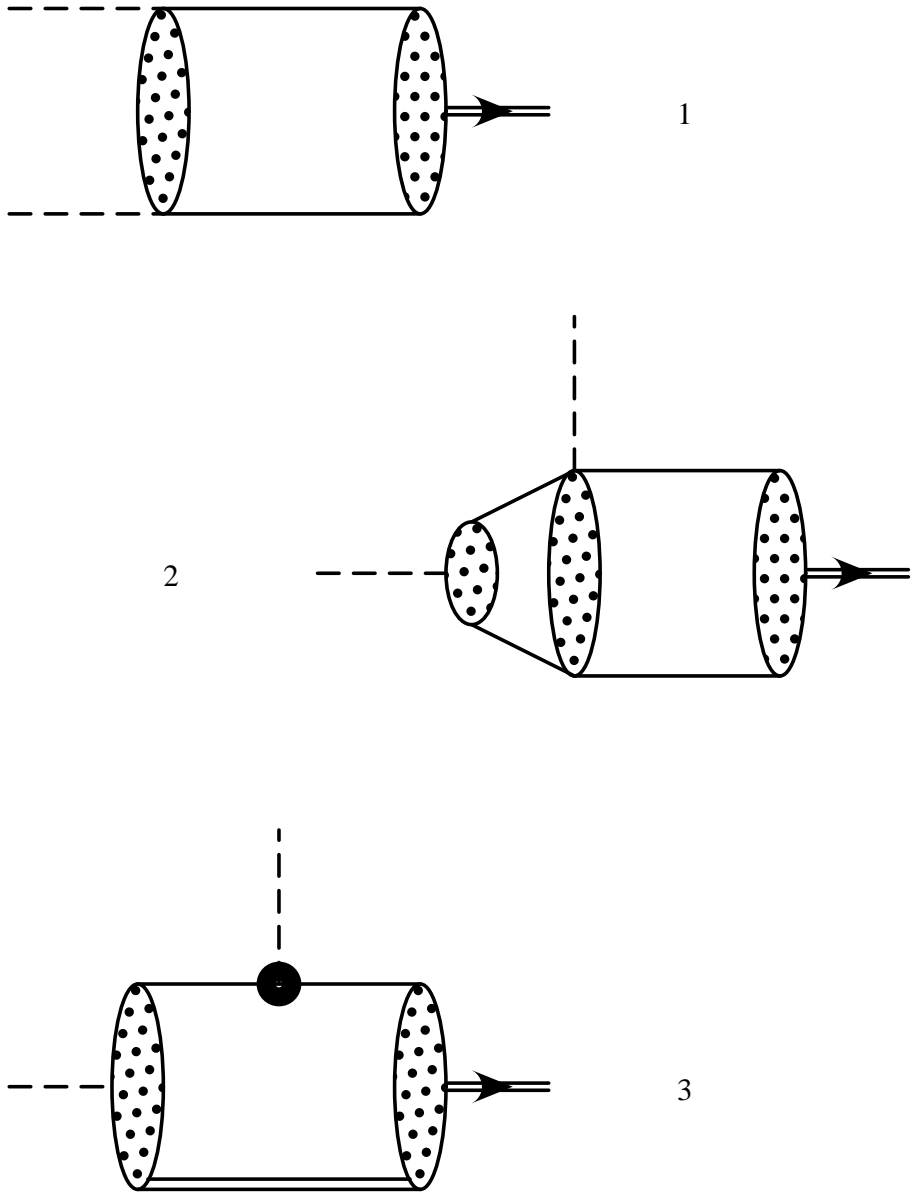}
\end{center}
\caption{The three regimes of that contribute to the
process we are considering
\rlabel{regimes}}
\end{figure}

 For large $Q^2$, the form factor $F_{\pi\gamma}(Q^2,\omega)$ can be
expressed as the convolution of the c\cf {} $C(x,\omega,Q,\mu)$
and the distribution function $\varphi(x,\mu^2)$
\bee
   F_{\pi\gamma}(Q^2,\omega)\ =\frac{N}{Q^2}\int^1_0 C(x,\omega,Q^2,\mu^2)
    \varphi(x,\mu^2)dx\ \equiv \frac{N}{Q^2}
    C(x,\omega,Q^2,\mu^2) \otimes \varphi(x,\mu^2) ,
 \rlabel{fpi}
\een
Here, $N= f_\pi(e^2_u-e^2_d)$ is a normalization factor,
$e_q$ are the chargers of the quarks,
$\mu^2$ is the renormalization mass, or the scale that separates
large from short distances,
and $f_\pi$ is the pion decay constant, normalized to
$f_\pi = 130$ MeV, see for example \rcite{PDG}.
The c\cf {} can be calculated in perturbation theory from the
hard parton subprocess $\gamma^{\ast}+\gamma^{\ast} \rightarrow
q\bar q$.  At present, the c\cf {} is known to leading twist to
one loop accuracy, and in the limit $\omega=1$ it has the simple form
\rcite{KMR}
\bee C(x,Q^2,\mu^2)\ =\ \frac{1}{2x}\left( 1 +
 \frac{\alfs}{4\pi} C_F\left[ \ln \left[\frac{2Q^2}{\mu^2}
  \right] (3+2\ln x)+\ln x^2-\ln\bar x- 9 \right] \right) \ +\{
  x\leftrightarrow {\bar x} \} .  \rlabel{1lcf}
  \een
 with $\bar x = 1-x$. The
distribution function can be determined by the moments \bee
f_\pi \int\limits^1_0 x^k\varphi(x,\mu^2)dx\ =\
\frac{i^k}{(Pn)^{k+1}}\ \langle0|\bar d\gamma_5(\hat{n})
(Dn)^ku|P\rangle
\rlabel{moments}
\een
where $n_{\mu}$ is a light-like vector, $\hat{n}= \gamma_{\mu}n^{\mu}$,
$D_{\mu}$ the covariant derivative and
$|P\rangle$ the $\pi^0$ meson state
with momentum $P$. By definition, $\varphi(x,\mu^2)$ is normalized to
one,
$
\int\limits^1_0 \varphi(x,\mu^2)dx\ =\ 1,$\ and G-parity implies
the relation
$\varphi(x)=\varphi(1-x)$.
Its $\mu^2$ dependence
is described by the evolution equation
\bee
\left[ \mu^2\frac{\partial}{\partial {\mu^2}}+
\beta(\alfs)\frac{\partial}{\partial \alfs}\right]
\varphi(x,\mu^2)\ =\int^1_0 V(x,y|\alfs )\varphi(y,\mu^2)dy
\rlabel{evol}
\een
%with some boundary condition
%%%%%%%%%%%%%%%%%%%%%%%%%%%%%%%%%%%%%%%
% change begins
%%%%%%%%%%%%%%%%%%%%%%%%%%%%%%%%%%%%%%%
with some initial condition
%%%%%%%%%%%%%%%%%%%%%%%%%%%%%%%%%%%%%%%
% change ends
%%%%%%%%%%%%%%%%%%%%%%%%%%%%%%%%%%%%%%%
\bee
        \varphi(x,\mu^2_0)\ =\ \varphi_0(x)
\een
The kernel $V(x,y|\alfs)$ is calculable  in pQCD, and can be
expanded in series of $\alfs$:
\be
V(x,y|\alfs)= \frac{\alfs}{4\pi}V^{(1)}(x,y)+
 \left(\frac{\alfs}{4\pi}\right)^2V^{(2)}(x,y)+ \cdots
\ee
It is known to two loop accuracy \rcite{MR}, \rcite{Sharmadi}.
In the one loop approximation \rcite{BL1}:
\beqn
V^{(1)}(x,y) &=&2C_F\left[\theta(y>x)\frac xy\left(1+\frac{1}{y-x}\right)+
(x\leftrightarrow \bar x,\ y\leftrightarrow \bar y) \right]_{+}\,
\rlabel{evolTREE} \\ \nonumber
\left[F(x,y)\right]_{+}&=&F(x,y)- \delta (x-y)\int^{1}_0F(t,y)dt.
\eeqn
Conformal symmetry, which at leading order manifests itself through
\bee
  y(1-y)V^{(1)}(x,y)=x(1-x)V^{(1)}(y,x),
\rlabel{1lsym}
\een
implies that the eigenfunctions that diagonalize the kernel
\rref{evolTREE}
are Gegenbauer polynomials multiplied by $x \bar x$.
One therefore expands the distribution function in this basis:
\bee
\varphi(x,\mu^2)\ =\ x\bar x
 \sum^\infty_{n=0}
b_{n}(\mu^2)\frac{2(3+4n)}{(1+n)(1+2n)}C^{3/2}_{2n}(1-2x)\ .
 \rlabel{1lsol}
\een
The coefficients $b_{n}(\mu^2)$ are given by
\bee
b_{n}(\mu^2)=b_{n}(\mu_0^2)\left[ \frac{\alfs(\mu^2)}
 {\alfs(\mu_0^2)}\right]^{ \gamma^{(1)}_n / 4 \pi b_0 }
 \rlabel{bn}
\een
Here $\gamma^{(1)}_n$ are the eigenvalues of $V^{(1)}(x,y)$,
or the one loop anomalous dimensions
of the multiplicatively renormalized operators:
\bee
\bar d\gamma_5(\hat{n})(n\partial)^{2n}_{+}
C^{3/2}_{2n}(nD_{-}/(n\partial)_{+})u
\rlabel{1lgo}
\een
where  $(n\partial)_{+}\equiv n\overrightarrow{\partial}+ n\overleftarrow{\partial}, \ \
 nD_{-}\equiv n\overrightarrow{D}- n\overleftarrow{D}$, see also
the operators in \rref{moments}.
The eigenvalues are given by:
\beqn
\gamma_n(\alfs)&=&\frac{\alfs}{4\pi}\gamma^{(1)}_n + \cdots\ ,\nonumber  \\
\gamma^{(1)}_n \ &=&\ C_F \left( 1-\frac{1}{(1+2n)(1+n)}
                           +4\sum^{2n+1}_{j=2}\frac{1}{j} \right) \ ,
\rlabel{1lad}
\eeqn
and $b_0 \ = \ (11/3N_c-2/3N_f)/4\pi$ is the first coefficient of the
QCD $\beta$~- function.
 To next-to-leading order, conformal symmetry is broken by
%%%%%%%%%%%%%%%%%%%%%%%%%%%%%%%%%%%%%%%%%%%%%%%%%%%%
% CHANGE BEGINS
%%%%%%%%%%%%%%%%%%%%%%%%%%%%%%%%%%%%%%%%%%%%%%%%%%%%
 renormalization of the coupling and by renormalization scheme
effects \rcite{BDFL},
 \rcite{Muller1}. The solution has a
%%%%%%%%%%%%%%%%%%%%%%%%%%%%%%%%%%%%%%%%%%%%%%%%%%%%
% CHANGE ENDS
%%%%%%%%%%%%%%%%%%%%%%%%%%%%%%%%%%%%%%%%%%%%%%%%%%%%
more complicated nondiagonal form, see for example
 \rcite{KMR}, \rcite{Muller2}.
Assuming that the expansion in Gegenbauer polynomials
(\ref{1lsol}) converges well, only the  first harmonics are
needed
 to obtain the model for the d\df {}. The coefficients
$b_{n}(\mu_0^2)$ in (\ref{bn}) should be extracted from the
%boundary function $\varphi_0(x)$, which is the low energy shape
%%%%%%%%%%%%%%%%%%%%%%%%%%%%%%%%%%%%%%%%%%%%%%%%%%%%%%%%%%%%%%%
%   CHANGE BEGINS
%%%%%%%%%%%%%%%%%%%%%%%%%%%%%%%%%%%%%%%%%%%%%%%%%%%%%%%%%%%%%%%
initial condition $\varphi_0(x)$, which is the low energy shape
%%%%%%%%%%%%%%%%%%%%%%%%%%%%%%%%%%%%%%%%%%%%%%%%%%%%%%%%%%%%%%%
%   CHANGE ENDS
%%%%%%%%%%%%%%%%%%%%%%%%%%%%%%%%%%%%%%%%%%%%%%%%%%%%%%%%%%%%%%%
of the distribution function, and can not be calculated within
pQCD. At present, there exist two popular models for
$\varphi_0(x)$.  These are $\varphi_{as}(x)=6x(1-x)$,
corresponding to the asymptotic d\df {} in the leading
logarithmic approximation, that is, only the first term in the expansion
(\ref{1lsol}), and $\varphi_{CZ}(x)=30x(1-x)(1-2x)^2$ which has
been proposed in \rcite{CZ}. In this model, the two first harmonics
are taken into account. The second coefficient
$b_1(\mu^2_0=1~{\rm GeV}^2)=3$ has been estimated using sum rules.
In this model, the second coefficient is large and
must be taken into account, while all the higher coefficients are
assumed  to be small and are neglected.
In \rcite{BF}, the convergence has claimed to be slow, and therefore
approximating the
distribution function to the first few terms by \rref{1lsol} might not
be justified.

 The relative contribution of all corrections
depends on the choice of the model. In particular, it has
been suggested in \rcite{MURA} that the discrepancy of the
predictions for $F_{\pi\gamma}(Q^2,\omega)$ within these models
will be large enough to allow for an experimental
discrimination.

\section{Calculation of the c\cf}

To sum all $(-b_0\alfs)^n$ contributions to the c\cf {}, we have to
calculate the c\cf {}  to leading order in the $1/N_f$ expansion,
and perform the replacement $2/3N_f \rightarrow 2/3N_f-11/3 N_c$
in the final
result, according to the prescription of NNA. In our case, we have
to calculate the one loop diagrams, but inserting a chain of $n$ fermion
bubbles. These diagrams yield factorialy growing contributions:
\bee
r_n \sim K(-\alfs b_0)^n n!n^b
\rlabel{assymtotic}
\een
The convergence radius of the series in \rref{assymtotic} is zero,
and in order to perform a ``summation'', Borel integral
techniques are used.
%%%%%%%%%%%%%%%%%%%%%%%%%%%%%%%%%%%%%%%%%%%%%%%%%%%%%%%%%%%
%  CHANGE BEGINS
%%%%%%%%%%%%%%%%%%%%%%%%%%%%%%%%%%%%%%%%%%%%%%%%%%%%%%%%%%%
The bad asymptotic behavior of \rref{assymtotic} will now manifest
itself through renormalon poles in the integrand of the Borel
integral, and a prescription has to be fixed to integrate over
these renormalon poles. In this work we are going to make use of
the principal value prescription.
The result will depend on how we have integrated over the poles,
being the ambiguity caused by the choice of a prescription known
to be suppressed by powers of $\Lambda^2/Q^2$. This ambiguity is 
usually referred to as renormalon ambiguity, and it can be cancelled
by taking the
contributions of higher twist operators into account: if the
renormalization of the lowest and higher twist contributions is
performed consistently, an ambiguity free result can be
obtained. 
%The prescription we are going to us is the principal
%value prescription.

In this work, we will evaluate the diagrams that contribute to the
c\cf {} to order $1/N_f$, see fig.\ref{1ldiagramms}.
The gluon line with a blob denotes the sum of all simple insertions
of fermion bubbles, fig.\ref{bubbles}. The Born contribution,
see fig.\ref{borndiagramms}, is well known since long ago.
We will use dimensional regularization to
regularize the ultraviolet divergences of these diagrams.  To
obtain the renormalized contribution from each diagram, first
the subdivergences, due to the fermion bubbles, and finally the
overall divergence of the whole diagram, have to be subtracted. This we
will do by following the technique presented in \rcite{BB2}.
Details on the calculations, and the separate contributions of
the diagrams can be found in Appendix A.  Here, we simply present
the final NNA results, together with some comments.  We will
present our results for the c\cf {} in the following way:
%%%%%%%%%%%%%%%%%%%%%%%%%%%%%%%%%%%%%%%%%%%%%%%%%%%%%%%%%%%
%  CHANGE ENDS
%%%%%%%%%%%%%%%%%%%%%%%%%%%%%%%%%%%%%%%%%%%%%%%%%%%%%%%%%%%

\begin{figure}
\begin{center}
\leavevmode\epsfxsize=12cm\epsfbox{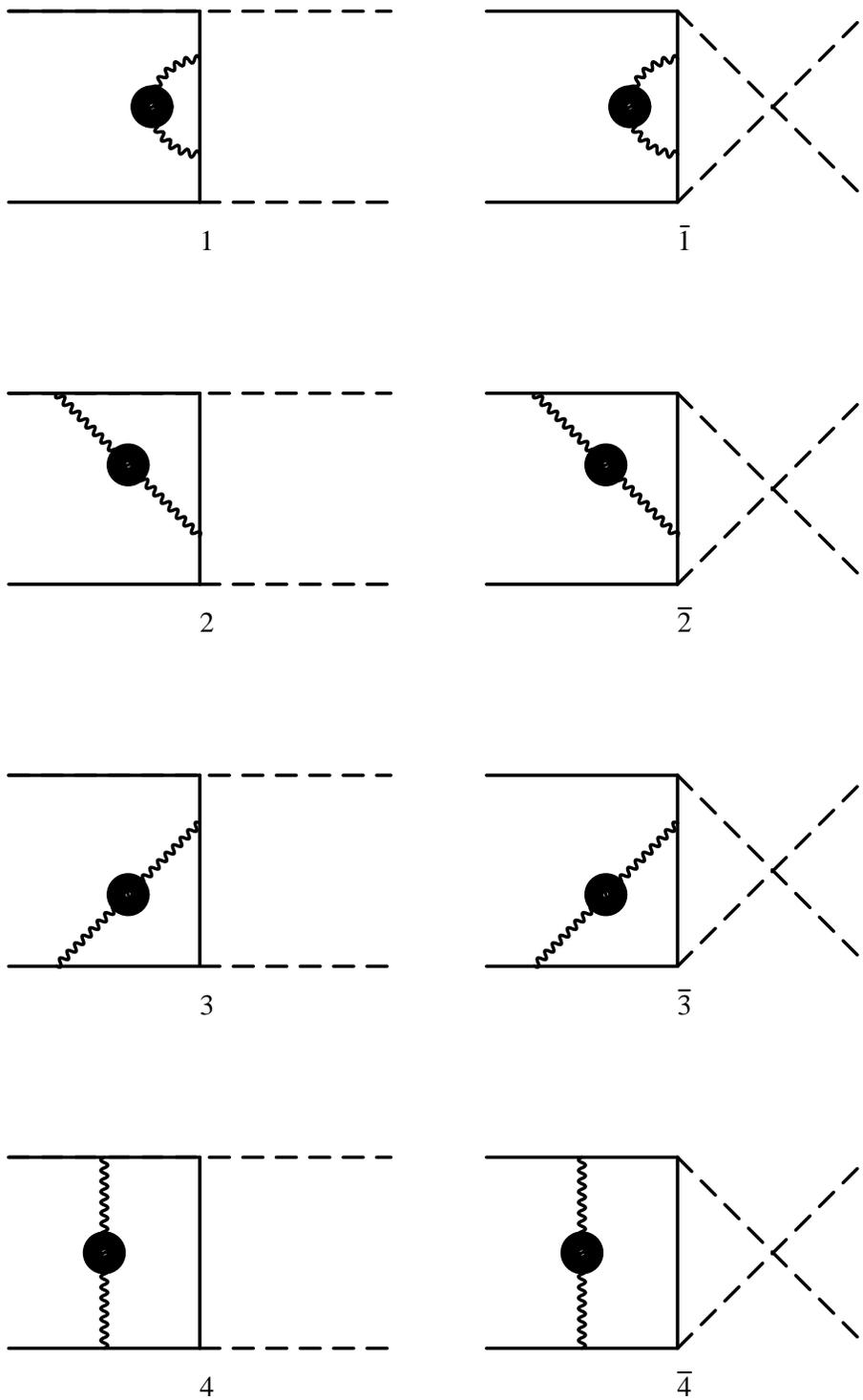}
\end{center}
\caption{ The diagrams that contribute to the coefficient
function in the leading order approximation
\rlabel{1ldiagramms}}
\end{figure}

\begin{figure}
\begin{center}
\leavevmode\epsfxsize=12cm\epsfbox{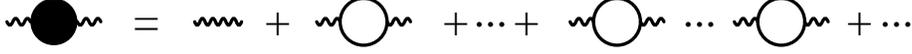}
\end{center}
\caption{ The full blob represents the sum of
all fermion bubble insertions
\rlabel{bubbles}}
\end{figure}

\bee
C(x,\omega, Q^2/\mu^2)\ =\ C_0(x,\omega)+C_0(1-x,\omega)+
C_1(x,\omega,Q^2/\mu^2)
  +C_1(1-x,\omega,Q^2/\mu^2)
 \rlabel{cf}
\een
Here $C_0$ is the leading order Born contribution
\bee
C_0(x,\omega) \ = \ \frac{1}{1-\omega+2x\omega}
\rlabel{Borncfw}
\een

\begin{figure}[h]
\begin{center}
\leavevmode\epsfxsize=12cm\epsfbox{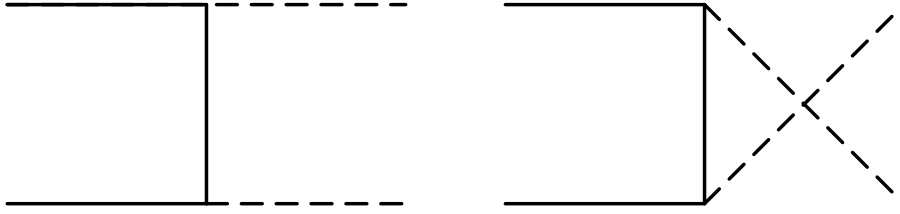}
\end{center}
\caption{ The two diagrams that contribute to the coefficient function in the
Born approximation.
\rlabel{borndiagramms}}
\end{figure}

In the next-to-leading order  contribution, $C_1$, a dependence
on $\alpha _s (\mu) $ is understood:
\beqn \nonumber
  C_1(x,\omega,Q^2/\mu^2) &=& -\frac{C_F}{4 \pi b_0}
C_0(x,\omega)\times
   \\ \nonumber
 & &\left\{ p.v. \int\limits^\infty_0 e^{-u/\alfs b_0} \frac{du}u
 \left[\left(\frac{\mu^2e^{C}}{Q^2}C_0(x,\omega)\right)^u \
 \frac{ 2\gamma(u|x,\omega)}{(1-u)(2-u)}\
 -\gamma(0|x,\omega)\right] \right. \nonumber \\ &+&
 \left.   \int\limits^{1+\alfs b_0}_1\ \frac{d\lambda}{1-\lambda}
 [G(\lambda|x,\omega)-G(1|x,\omega)]\right\}\ ,\rlabel{firC1}\\
\gamma(u|x,\omega) &=&
3+{}_2F_1 \left[{1,1 \atop 2+u}\left|
\frac{2\omega\bar x}{1+\omega}\right.\right]
\left[\frac{2-u}{2\omega C_0(x,\omega)}-
\frac2{1+u}\right] \nonumber \\
&-&
_2F_1\left[{1,1 \atop 2+u}\left| -\frac{2\omega x}{1-\omega}
\right. \right]\left[
\frac{2-u}{2\omega C_0(x,\omega)}+\frac2{1+u}\right]\
\eeqn
\beqn
G(\lambda|x,\omega)&=&
\frac13 \frac{\Gamma (2\lambda+2)}{\Gamma(2-\lambda)\Gamma^3(1+\lambda )}
\left\{ \frac{2+2\lambda-\lambda^2}{1+\lambda}
+ {}_2F_1 \left[{1,\lambda \atop 1+\lambda}\left|
\frac{2\omega\bar x }{1+\omega}\right. \right]
\left[\frac{\lambda}{2\omega C_0(x,\omega)}-1 \right]\
\right. \nonumber \\
&-&\left. {}_2F_1 \left[{1,\lambda \atop 1+\lambda}\left|-
\frac{2\omega x}{1-\omega}\right.\right]
\left[\frac{\lambda}{2\omega C_0(x,\omega)}+1 \right]\
\right\} \ , \rlabel{cf1}
\eeqn
where
$p.v.$ means that we use the principal value prescription to integrate
 over the $IR$-renormalon poles
at $u=1$ and $u=2$. $C$ parametrizes the renormalization scheme. In the
$MS$ scheme,
 $C=\ln(4\pi)-\gamma_E+5/3$
\footnote{Notice that our $C$, and the $C$ used in \rcite{BB2,BBB}
have different global signs.}
  ($\gamma_E$ is the Euler constant) and
in the $\overline{MS}$-scheme, $C=5/3$.
The fact that we have to fix a prescription to integrate over the
poles at $u=1$ and $u=2$ induces an ambiguity  $\delta C_1(x,\omega)$
%%%%%%%%%%%%%%%%%%%%%%%%%%%%%%%%%%%%%%%%%%%%%%%%%%
% CHANGE BEGINS
%%%%%%%%%%%%%%%%%%%%%%%%%%%%%%%%%%%%%%%%%%%%%%%%%%
in the c\cf {}. It is well known that the ambiguity induced by a 
singularity 
%simple pole singularity 
%%%%%%%%%%%%%%%%%%%%%%%%%%%%%%%%%%%%%%%%%%%%%%%%%%
% CHANGE ENDS
%%%%%%%%%%%%%%%%%%%%%%%%%%%%%%%%%%%%%%%%%%%%%%%%%%
at $u=u_0$ is power suppressed by $(\Lambda^2/ \mu ^2)^{u_0}$.
In our case, for $\mu^2=Q^2$, the ambiguity of the coefficient
function will read
\bee
\delta C_1(x,\omega) = \pm \Delta _2 (x,\omega)\frac{\Lambda
^2}{Q^2} \pm \Delta _4 (x,\omega) \left( \frac{\Lambda ^2}{Q^2}
\right) ^2
\een
where $\Delta _2 (x,\omega)$ and $\Delta _4
(x,\omega)$ are the residues of the poles of the Borel integral,
\rref{firC1}, at $u=1$ and $u=2$ respectively. The $IR$-renormalon
ambiguity $\delta C_1(x,\omega)$ has to be canceled exactly by
another ambiguity, the $UV$-renormalon ambiguity of the matrix
elements of higher-twist operators.  This means that higher
orders in perturbation theory and higher twist contributions are
inseparable.  This fact can be used to obtain information on
higher twist effects.  According to the assumption of ultraviolet
dominance, the full higher twist contributions are proportional to
the $UV-$renormalon contributions, and the
entire higher twist contribution can be included in the following way:
 \bee F_{\pi\gamma}(Q^2,\omega)\
= \frac{N}{Q^2}\int^1_0 \left\{ C(x,\omega,Q^2) +
N_2\Delta_2(x,\omega)\frac{\Lambda^2}{Q^2}
+ N_4\Delta_4(x,\omega)\frac{\Lambda^4}{Q^4}+ \cdots  \right\}
    \varphi(x,Q^2)dx\
\een
The dots denote higher power contributions. Within the assumption of
ultraviolet dominance, \rcite{NNAbraun} the complete dependence on the
kinematic variables $x,\omega$ is fixed by the calculable functions
$\Delta_2(x,\omega),\Delta_4(x,\omega)$. The constants $N_i$ and
their sign have to
be fixed from experiment and it seems reasonable to expect
their values to be of order one. In what follows, we set
these constants to plus minus one and use this range as an
estimate of the higher-twist effects. It should be kept in mind that
this is a (perhaps raw) estimate. In Deep Inelastic Scattering, $N_i \sim 2$
\rcite{DIS}, in support of our estimation.

%%%%%%%%%%%%%%%%%%%%%%%%%%%%%%%%%%%%%%%%%%%%%%%%%%%%%%%%%%%%%%%%%%%
%  CHANGE BEGINS
%%%%%%%%%%%%%%%%%%%%%%%%%%%%%%%%%%%%%%%%%%%%%%%%%%%%%%%%%%%%%%%%%%%

We now turn to the region that is accessible to experiment, that is,
%$\omega \rightarrow 1$.
$\omega = 1$.
Keeping only leading corrections, we obtain
\begin{eqnarray}
%C_0(x,\omega) &=& \frac1{2x}\ + \ O(1-\omega)\ \rlabel{Borncf}  ,\\
C_0(x,1) &=& \frac1{2x}\ \rlabel{Borncf} ,\\
 C_1(x,1,Q^2/\mu^2)&\equiv& C_1(x,Q^2/\mu^2), \nonumber \\
% C_1(x,\omega,Q^2/\mu^2)&=& C_1(x,Q^2/\mu^2)\ +\
% (1-\omega)\ln(1-\omega)C^{\prime}_1(x,Q^2/\mu^2)+
% O(1-\omega)\ \nonumber , \\
 C_1(x,Q^2/\mu^2)&=&
\frac{(-)C_F}{4 \pi b_0}\frac1{2x}\ \left\{
p.v. \int\limits^\infty_0
e^{-u/\alfs b_0} \frac{du}u \left[
\frac{2}{(1-u)(2-u)}\left(\frac{e^{C}\mu^2}{2xQ^2}
\right)^{u}\gamma(u,x)-\gamma(0,x)\right] \right. \nonumber \\
&+&\left. \int\limits^{1+\alfs b_0}_1
\frac{d\lambda}{1-\lambda}[G(\lambda,x) -G(1,x)]
\right\} \rlabel{NNAcf}  \\
\gamma(u,x) &=& 3+\,_2F_1\left.\left[
{1,1 \atop 2+u}\right|1-x \right]\left\{
x(2-u)-\frac2{1+u}\right\}\ , \nonumber  \\ \nonumber
\\
G(\lambda,x) &=&
\frac13 \frac{\Gamma (2\lambda+2)}{\Gamma(2-\lambda)\Gamma^3(1+\lambda )}
 \left\{ \frac{2+2\lambda-\lambda^2}{1+\lambda}+(\lambda x-1)
_2F_1\left.\left[ {1,\lambda \atop 1+\lambda}\right| 1-x\right] \right\}
 \nonumber
\end{eqnarray}
%\begin{eqnarray*}
% C^{\prime}_1(x,Q^2/\mu^2)&=&\frac{(-)C_F}{4 \pi b_0}
% \left(\frac1{2x}\right)^2 \times \\
%& &%\left\{
% p.v.\int\limits^\infty_0 e^{-u/\alfs b_0} \frac{du}u
% \left[
% \left(\frac{e^{C}\mu^2}{2xQ^2}\right)^{u}
% \frac{2(1+u)}{(1-u)(2-u)}\left\{x(2-u)+\frac2{1+u}\right\}-2(1+x)
% \right]
% \\ \nonumber
%\end{eqnarray*}

%%%%%%%%%%%%%%%%%%%%%%%%%%%%%%%%%%%%%%%%%%%%%%%%%%%%%%%%%%%%%%%%%%%
%  CHANGE ENDS
%%%%%%%%%%%%%%%%%%%%%%%%%%%%%%%%%%%%%%%%%%%%%%%%%%%%%%%%%%%%%%%%%%%
We have checked for $\omega =\ 1$ that in the one loop limit
these formulae are in agreement
with \rref{1lcf}.
Consider  $C_1(x,Q^2/\mu^2)$. Using the simple identities
\beqn
\frac1{u(1-u)(2-u)} &=& \frac1{2u} + \frac1{(1-u)}- \frac1{2(2-u)}\ , \\
\exp\{-u/\alfs (\mu^2)b_0\}t^{u} &=& \exp\{-u/\alfs (\mu^2/t)b_0\}
\eeqn
we can rewrite the Borel integral in \rref{NNAcf} in the following way
\beqn
& &p.v. \int\limits^\infty_0
e^{-u/\alfs b_0} \frac{du}u \left[
\frac{2}{(1-u)(2-u)}\left(\frac{e^{C}\mu^2}{2xQ^2}
\right)^{u}\gamma(u,x)-\gamma(0,x)\right] = \\ \nonumber
& &p.v. \int\limits^\infty_0
\exp\left\{\frac{-u}{\alfs (2xQ^2e^{-C}) b_0}\right\}
\left[
\frac{\gamma(u,x)-\gamma(0,x)}{u}+
2\frac{\gamma(u,x)}{1-u}-
\frac{\gamma(u,x)}{(2-u)}
\right] du \\ \nonumber
& &+
\gamma(0,x) \log { \alfs (2xQ^2e^{-C}) \over \alfs( \mu ^2 ) }
\eeqn
It is now easy to see that the integral diverges for
$x<e^C \Lambda ^2 /2Q^2$ due to the Landau pole in the running coupling.
This effect arises because our effective expansion parameter
is $\Lambda^2/(xQ^2)$.
This means that for small values of $x$, the entire power
expansion in $1/Q^2$ needs to be resummed.
A similar situation has been discussed recently in \rcite{BBM}.
It has been
shown that this resummation leads to new power corrections.
It seems reasonable to assume that
in our case these new power corrections are related to the fact
%%%%%%%%%%%%%%%%%%%%%%%%%%%%%%%%%%%%%%%%%%%%%%%%%%%%%%%%%
%   CHANGE BEGINS
%%%%%%%%%%%%%%%%%%%%%%%%%%%%%%%%%%%%%%%%%%%%%%%%%%%%%%%%%
that for $\omega=1$, standard factorization breaks down for the
higher twist contributions, and new regimes
%%%%%%%%%%%%%%%%%%%%%%%%%%%%%%%%%%%%%%%%%%%%%%%%%%%%%%%%%
%   CHANGE ENDS
%%%%%%%%%%%%%%%%%%%%%%%%%%%%%%%%%%%%%%%%%%%%%%%%%%%%%%%%%
have to be taken into account, see discussion in Section 1.

We will overcome this small $x$ problem following the approach of
\rcite{BBM}.
We will first convolute the coefficient function, \rref{NNAcf},
with the wave function, and
afterwards do the Borel integral. The result will of
course depend on the distribution function we have convoluted with,
but in general, we will encounter a new singularity structure. For
the asymptotic wave function, $\varphi _{as}(x) =6x(1-x)$,
for example, we will find singularities for all positive
integers $u$, simple poles for $u > 2$, and double poles for
$u=1$ and $u = 2$. A very interesting situation arises with the
NNA asymptotic distribution function, see next section for
details. Here, the new singularities will be located at
$u = b_0\alfs(Q^2)+m$, with $m=1,2,\ldots$. Despite the position of
these new poles, the power corrections they induce are integer powers
of $\Lambda ^2/Q^2$:

\be
\left( \Lambda ^2 \over Q^2 \right)^{m+b_0\alfs(Q^2)} =
\left( \Lambda ^2 \over Q^2 \right)^
{m+ 1/ \log ( Q^2 / \Lambda ^2) } = e^{-1}
\left( \Lambda ^2 \over Q^2 \right)^m
\ee
and our leading order result is in agreement with the predictions
of the power counting rules.

\section{The distribution function}

\subsection{The distribution function in the NNA approximation}
The pion distribution function is a phenomenological model function,
and information about its shape should be taken either from experiment,
or from nonperturbative calculations. In perturbative QCD it is only possible
to predict its evolution with $\mu ^2$ using the evolution equation
(\ref{evol}). At the one loop level, conformal symmetry allows
to find a basis of multiplicatively renormalized operators,
\rref{1lgo}. It follows that the
eigenfunctions that diagonalize the evolution kernel, \rref{evolTREE},
are Gegenbauer polynomials multiplied by $x \bar x$. This suggests to
expand $\varphi (x)$ in terms of these eigenfunctions, \rref{1lsol}.
Since for $n>0$,
$\gamma_n^{(1)}>0$ and $\gamma_0=0$, see \rref{1lad}, only the lowest
harmonic in (\ref{1lsol}) survives in the limit
$\mu^2 \rightarrow \infty$. This leads to one of the most popular models
for the distribution function, the asymptotic distribution
function, $\varphi_{as}(x)=6x(1-x)$.
The two loop  corrections break conformal invariance. The operators
\rref{1lsol} get mixed, and the evolution kernel can no longer be
diagonalized with the basis of Gegenbauer polynomials. As a
consequence, the evolution of the distribution function is now
much more complicated than in the previous case, see for example
\rcite{KMR}, \rcite{Muller2}.
Below, we present the analysis of the distribution function within the NNA
approximation.

Our starting point is the lowest order NNA evolution equation:
\begin{equation}
\left[ \mu^2\frac{\partial}{\partial \mu^2}
-b_0 \alfs ^2\frac{\partial}{\partial \alfs} \right]
\varphi(x,\mu^2)\ =\int^1_0 V[x,y|\alfs(\mu^2)]\varphi(y,\mu^2)dy
\rlabel{NNAevol}
\end{equation}
The NNA evolution kernel  reads
 \begin{eqnarray*}
V[x,y|\alfs(\mu^2)]\equiv V_{\alpha}(x,y)&=&\frac{\alfs}{4\pi}C_F\frac\alpha3\
\frac{\Gamma(2\alpha+2)}{\Gamma(2-\alpha)\Gamma^3(1+
\alpha)}\times \\
&&  \left[\Theta(y>x)\left(\frac
xy\right)^\alpha\left( \alpha+\frac1{y-x}\right)+\{x\leftrightarrow\bar x,
y\leftrightarrow \bar y\}\right]_+\ .
\end{eqnarray*}
Here, we have introduced the notation $\bar x\equiv 1-x,\
\bar y\equiv 1-y\ , \  \alpha\equiv 1+\alfs b_0$ and as usual
$[F(x,y)]_+\ =\ F(x,y)-\delta(x-y)\int^1_0\ dt\,F(t,y) $.
We have obtained this kernel combining the techniques presented in
\rcite{MR} and \rcite{BB2}.

\begin{figure}
\begin{center}
\leavevmode\epsfxsize=12cm\epsfbox{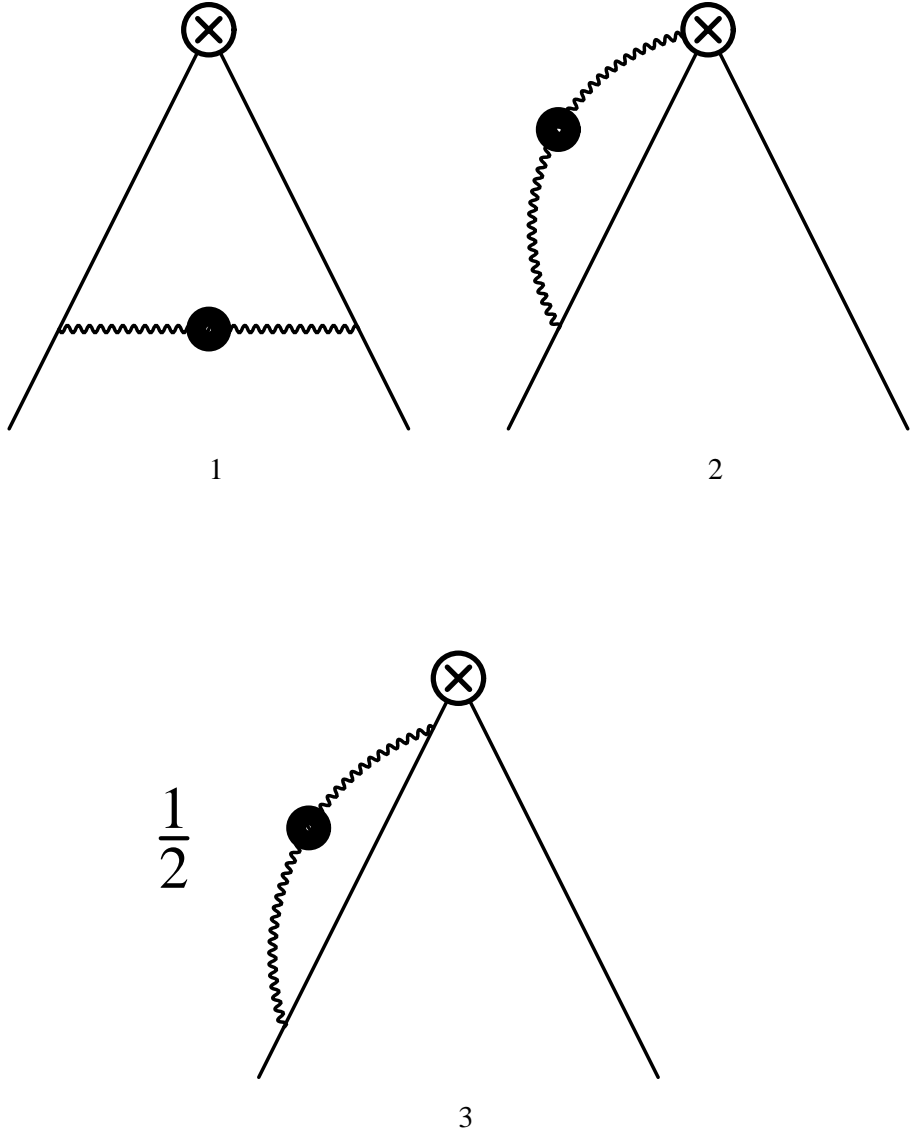}
\end{center}
\caption{The diagrams that contribute to the evolution kernel in the
leading order approximation. For diagrams (2) and (3), the mirrors conjugate
diagrams have to be added.
\rlabel{kerneldiagr}}
\end{figure}

 The relevant diagrams are shown in fig.\ref{kerneldiagr}
\footnote{ As a byproduct of our calculation we also obtained the
evolution kernel $P(Z)$ for the forward case. We agree with the result
presented in \rcite{Mikha} }.
This kernel, as in the one loop case,  \rref{1lsym}
becomes symmetric after multiplication by $(y\bar y)^\alpha$:
$$ (y\bar y)^\alpha V_{\alpha}(x,y)\ =\ W(y,x)\ =\ W(x,y)\ . $$
This fact allows us to obtain the eigenfunctions and eigenvalue of
the kernel $V_{\alpha}(x,y)$ 
%%%%%%%%%%%%%%%%%%%%%%%%%%%%%%%%%%%%%%%%%%%%%%%%%%%%%%%%
% CHANGE BEGINS
%%%%%%%%%%%%%%%%%%%%%%%%%%%%%%%%%%%%%%%%%%%%%%%%%%%%%%%%
\footnote{After finishing our calculations,
Mikhailov published a work \rcite{Mikha},
where this possibility is also discussed. } .
%%%%%%%%%%%%%%%%%%%%%%%%%%%%%%%%%%%%%%%%%%%%%%%%%%%%%%%%
% CHANGE ENDS
%%%%%%%%%%%%%%%%%%%%%%%%%%%%%%%%%%%%%%%%%%%%%%%%%%%%%%%%
 \begin{eqnarray} \int^1_0
V_{\alpha}(x,y)\bar\varphi_{n}(y,\mu^2)&=&-\gamma_n(\alfs)
\bar\varphi_{n}(x,\mu^2)\ , \\
\bar\varphi_{n}(x,\mu^2)&=&
(x\bar x)^{\alpha}\varphi_{n}(x,\mu^2)A_{n}(\alfs)\ ,
\rlabel{NNAegf} \\
\varphi_{n}(x,\mu^2)&=& C^{1/2+\alpha}_{2n}(1-2x)\ , \\
 A_{n}(\alfs) &=&
\frac{\Gamma(1+2\alpha)}{\Gamma(\alpha)\Gamma(1+\alpha)}
\frac{(2n)!}{(2\alpha)_{2n}}
\frac{(1+2\alpha+4n)}{\alpha+n} \ \rlabel{An}, \\
\int^1_0\varphi_{n}\bar\varphi_{k}dx &=& \delta_{kn}\ ,
\end{eqnarray}
The eigenvalues $\gamma_n(\alfs)$ read:
\begin{eqnarray}
\gamma_n(\alfs)
&=&\frac{\alfs}{4\pi}C_F \frac{\alpha^2\Gamma(2\alpha+2)}{3(1+\alpha)\
\Gamma(1+\alpha)^3\Gamma(2-\alpha)}\left\{1-\frac{\alpha(\alpha+1)}{
(\alpha+2n)(\alpha+1+2n)} \right. \nonumber \\
&+& \left. \frac{2(1+\alpha)}\alpha \left[
\psi(1+\alpha+2n)-\psi(1+\alpha)\right]
\right\}.
\rlabel{NNAad}
\end{eqnarray}
Here, $\psi (z)$ is given by $\psi(z)=\frac{d}{d z}\ln \Gamma(z)$.
Notice that,  as in the one loop case, quark current
conservation implies $\gamma_0(\alfs)=0$.
From \rref{NNAegf} we see that in our case  it is natural to
expand the d\df {} in $C_{2n}^{3/2+\alpha _s b_0}$
Gegenbauer polynomials:
\begin{equation}
\varphi(x,\mu^2)\ =\ (x \bar
x)^{1+\alfs b_0} \sum^\infty_{n=0}
b_{n}(\mu^2)A_{n}(\alfs)C^{3/2+\alfs b_0}_{2n}(1-2x)\ ,
\rlabel{NNAsol}
\end{equation}
compare with \rref{1lsol}.
Substituting (\ref{NNAsol}) in the evolution equation (\ref{NNAevol})
and using orthogonality of
$\bar\varphi_{n}$ and $\varphi_{k}$, we obtain the following
equation for the
moments $b_{k}(\mu^2)$:
\bee
 \left[ \mu^2\frac{\partial}{\partial \mu^2}-
 b_0\alfs^2\frac{\partial}{\partial \alfs}\right]
 b_{k}(\mu^2)=-b_{k}(\mu^2)\gamma_{k}(\alfs)-
 \sum^{k}_{\ell=0} C_{k \ell}(\alfs)b_{\ell}(\mu^2)\ .
\rlabel{mixed}
\een
where we have introduced
the mixing matrix $C_{k \ell}(\alfs)$, which arises due
to the fact that now the eigenfunctions depend on $\alfs(\mu^2)$.
Technical details can be found in  Appendix~B.
In \rref{mixed}, $b_0$ is the first coefficient of the $\beta$ function,
and should not be confused with the moment $b_0(\mu)$, where an
explicit $\mu$ dependence is indicated.
 We obtain a triangular system of linear differential equations for
the coefficients $b_{p}(\mu^2), \ p=0,1,\ldots,k$.
Introducing the  vector
\be
 {\bf B}\ =\ \left(\begin{array}{l} b_0(\mu^2) \\ b_1(\mu^2) \\
\vdots \\ b_{k}(\mu^2) \\ \vdots   \end{array} \right)
\rlabel{vecB}
\ee
our equations can be written in the following matrix form:
\bee
\left[ \mu^2\frac{\partial}{\partial \mu^2}
-b_0\alfs^2\frac{\partial}{\partial \alfs}\right]
{\bf B}(\mu^2)\ =\ - (\hat\gamma_D+\hat C){\bf B}(\mu^2)\ .
\rlabel{Mevol}
\een
Here, $\hat\gamma_D$ is a diagonal matrix, built of the
eigenvalues (\ref{NNAad}), and $\hat C$ is the triangular
mixing matrix of \rref{mixed}.
The first $N$ elements of \rref{vecB} can be obtained
exactly for arbitrary $N$. One first notices that \rref{mixed}
implies that $b_0(\mu^2)$ is constant because $C_{00} =\gamma_0=0$
(due to conservation of the axial current).
For $b_1(\mu^2)$,
\rref{mixed} implies

\be \left[ \mu^2 {\partial
\over \partial \mu^2} -b_0 \alfs ^2 {\partial \over \partial
\alfs} \right] b_1(\mu^2) = - [\gamma_1 (\alfs) + C_{11}(\alfs)
] b_1(\mu^2) -C_{10} (\alfs) b_0(\mu^2), \rlabel{b1ECUA}
\ee %
which can be solved using well known techniques. This procedure can now
be repeated for $b_2(\mu^2)$, $b_3(\mu^2)$, $\ldots$.

The general solution to \rref{mixed} is given by
\bee
 {\bf B}(\mu^2)\ =\ \hat U(\mu^2,\mu^2_0){\bf B}(\mu^2_0)
\een
with the evolution matrix
\begin{eqnarray}
\hat U(\mu^2,\mu^2_0)&=&P_a\exp\left\{\int\limits^{a(\mu^2)}_{
a(\mu^2_0)} \frac{da'}{b_0 {a'}^2}[\hat\gamma(a')+\hat
C(a')]\right\}\\ \hat U(\mu^2_0,\mu^2_0) &=& \hat 1\ .
\rlabel{NNAgsol}
\end{eqnarray}
It follows that
\bee
 b_{k}(\mu^2)\ =\ \sum^k_{k'=0} \left\{
\hat U(\mu^2,\mu^2_0)\right\}_{kk'}b_{k'}(\mu^2_0),
\een
and we obtain the following expression for $\varphi(x,\mu^2)$:
\bee
 \varphi(x,\mu^2)=(x\bar x)^{1+\alfs b_0}\sum^\infty_{k=0}\left\{\hat
   U(\mu^2,\mu^2_0)\right\}_{kk'}b_{k'}(\mu^2_0)A_{k}(\alfs)
   C^{3/2+\alfs b_0}_{2k}(1-2x)\ ,
\een
where we sum for $k'\le k$.

We have found numerically that the matrix elements of
the nondiagonal part of the
mixing matrix $\hat C$ are much  smaller than the diagonal matrix
elements of the sum $\hat\gamma+\hat C$ for all resonable values
of $\mu^2$.
 We plot the matrix elements for
$\mu^2=5$ GeV$^2$ in  fig.\ref{mix}.a and  fig.\ref{mix}.b.
The diagonal elements are clearly bigger than the non diagonal ones.
This happens for the whole range of relevant $\mu ^2$.

\begin{figure}[t]
\begin{tabular}{cc}
\mbox{${{\leavevmode\epsfxsize=8cm\epsfbox{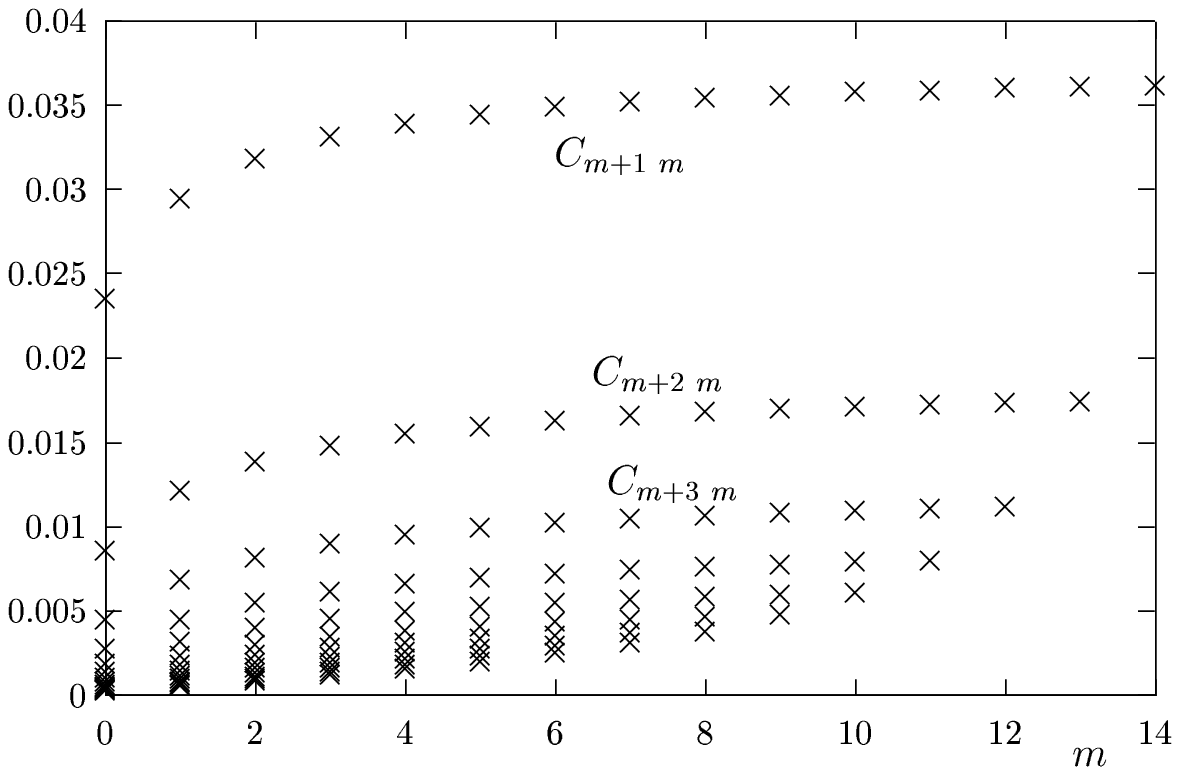}}}$} &
\mbox{ $
{\leavevmode\epsfxsize=8cm\epsfbox{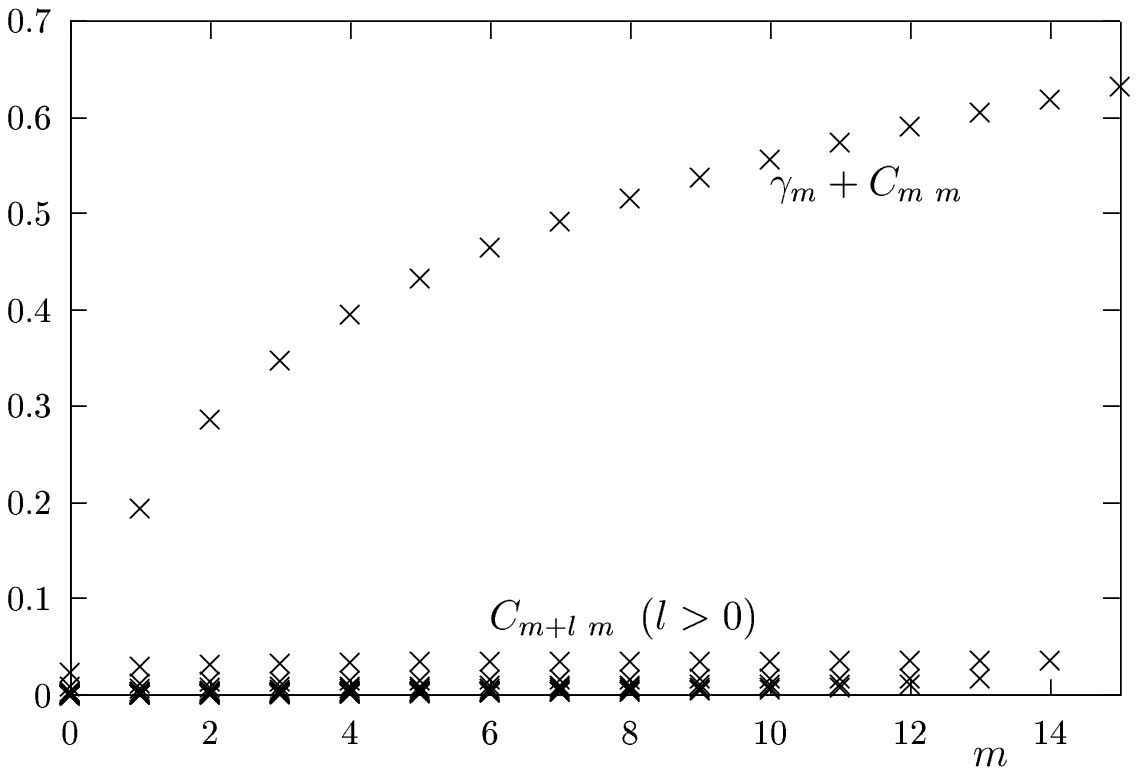}}$
} \\
\mbox{\ref{mix}.a} & \mbox{\ref{mix}.b} \\
\mbox{$\ $} & \mbox{$\ $} \\
\mbox{${{\leavevmode\epsfxsize=8cm\epsfbox{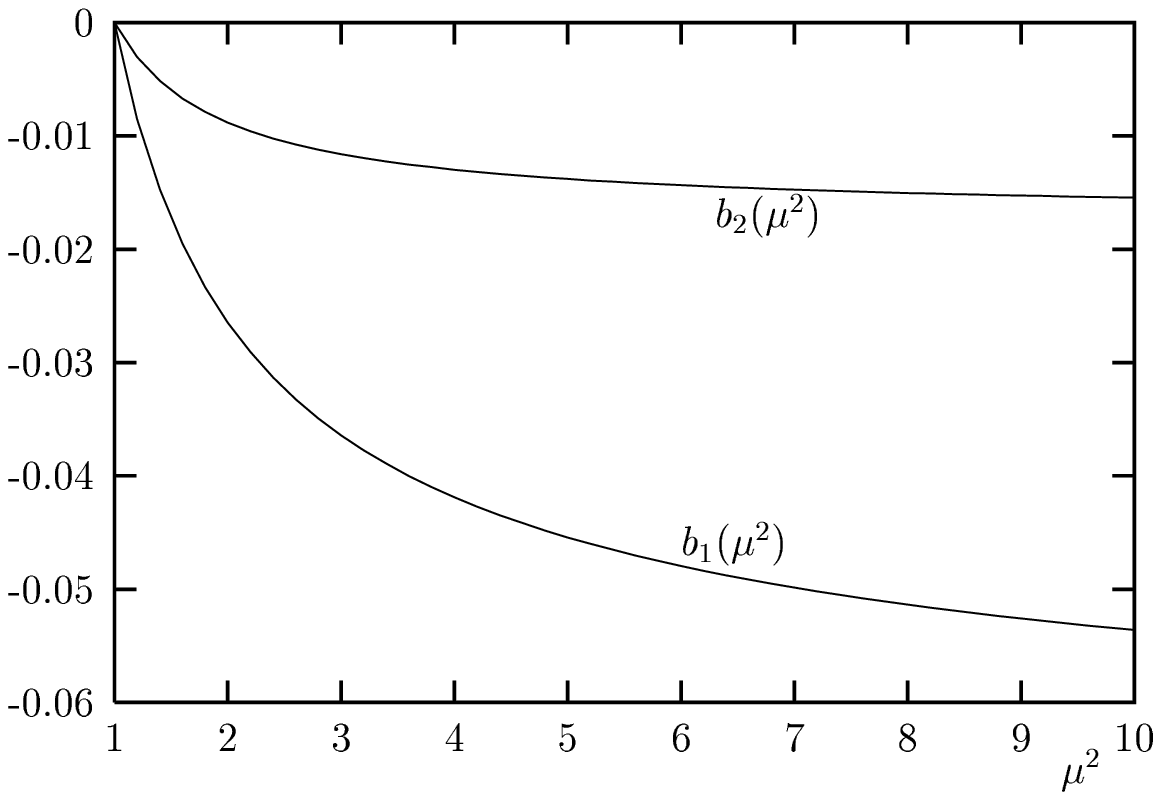}}}$} &
\mbox{ $
{\leavevmode\epsfxsize=8cm\epsfbox{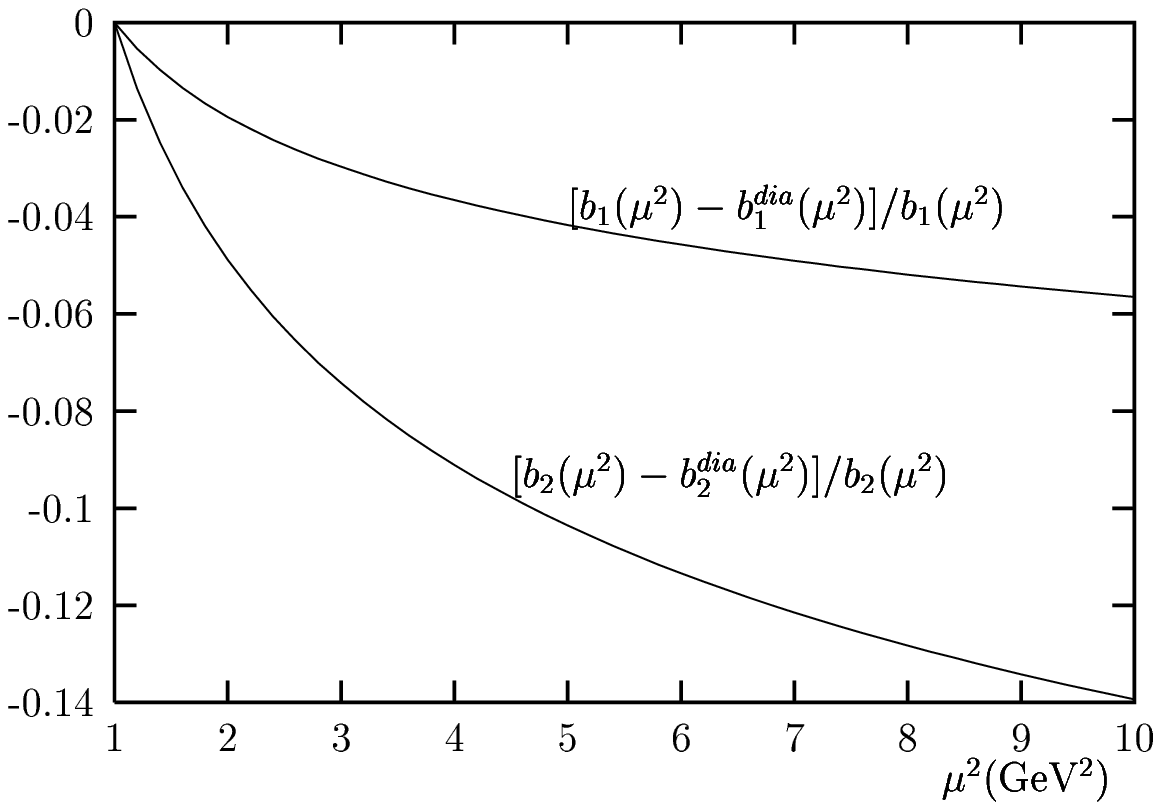}}$
} \\
\mbox{\ref{mix}.c} & \mbox{\ref{mix}.d}
\end{tabular}
            \caption{
In \ref{mix}.a we show
the nondiagonal elements $C_{k\ l}$ for
$k=0,1,\cdots,15$,  $l=0,1,2,\cdots,k-1$ and $\mu^2 = 5$  GeV$^2$.
In \ref{mix}.b we compare  the nondiagonal elements, \ref{mix}.a,
with the
diagonal ones, $\gamma_m+ C_{mm}, \ m\leq 15$.
The effect of the nondiagonal terms on the evolution
of $b_1(\mu^2),b_2(\mu^2)$ is displayed for
$(b_0,b_1,b_2) (1$ GeV$^2)= (1,0,0)$
(\ref{mix}.c), and for $(b_0,b_1,b_2) (1$ GeV$^2)= (1,1.7,1.6)$
(\ref{mix}.d).
 \rlabel{mix}  }
 \end{figure}

We can therefore
solve equation \rref{Mevol} by iterations with respect to the
nondiagonal part $\hat C$.
At leading order, neglecting the nondiagonal part of the mixing
matrix, all coefficients $b_{k}(\mu^2)$ renormalize multiplicatively,
and the distribution function has the following form:
\beqn
 \varphi(x,\mu^2)&=&(x\bar x)^{1+b_0\alfs}\sum^\infty_{k=0}
           b_{k}(\mu^2)A_{k}(\alfs)
           C^{3/2+\alfs b_0}_{2k}(1-2x)\ ,
 \rlabel{NNAdf}   \\
 b_{k}(\mu^2)&=&
 b_{k}(\mu_0^2)\exp
 \left\{
 \int_{\alfs(\mu^2_0)}^{\alfs(\mu^2)} \frac{\gamma_k(x)+
 C_{kk}(x)}{b_0x^2}dx
 \right\} ,
 \rlabel{bUk}
\eeqn
where
\bee
C_{kk}(\alfs)= (\alfs b_0)^2
\left\{\psi\left(\frac12+\alpha+2k\right)
   -\psi\left(\alpha+\frac12\right)\right\}.
\een

\subsection{Models for the distribution function}
We have already mentioned that the distribution functions is not 
predicted
by perturbative QCD. In this subsection we discuss some models for
the distribution function.
For any physical distribution function, the first coefficient
$b_0(\mu^2)$ of the expansion \rref{NNAdf} is 1, as it follows from
normalization
\be
\int\limits^1_0 \varphi(x,\mu^2)dx\ =\ 1.
\ee
To extract the  coefficients $b_{k}(\mu_0^2)$ for  $k>0$, some information
about the low energy shape of the distribution function is needed. At
present, some sum rule estimates are available. The second moment
\bee 
\int_0^1 (2x-1)^2 \varphi(x,\mu^2 _0 )dx =0.35 \quad (\mu^2_0
=1\ \rm{GeV}^2)
 \rlabel{mom1} \een has been estimated in
\rcite{CZ,BF}, and the following estimation for $\varphi(1/2)$
can be found in \rcite{BF}:
\bee \varphi(1/2,\mu^2_0)= 1.2\pm
0.3 \rlabel{1/2}
\een
For the second moment, the asymptotic d\df
{} predicts
 \bee \int_0^1 \varphi_{as}(x)(2x-1)^2=0.2 , \rlabel{mom2}
\een
and for $x = 1/2$, its prediction is $\varphi_{as}(1/2)=3/2$.

The estimation \rref{mom1} would imply $b_1 (\mu _0 ^2) = 1.1$
in the one loop expansion \rref{1lsol}, and $b_1 (\mu _0 ^2) = 1.7$ in
the NNA expansion \rref{NNAdf}. The coefficient in the NNA expansion
is $45 \%$ bigger than the 1 loop coefficient. To be consistent with the
central value of \rref{1/2}, we need at least the third coefficient
of the expansion of the distribution functions, that is,
\bee
\varphi(x)=[x(1-x)]^{1+b_0\alfs}\left( A_0(\alfs)+
b_1A_1C_2^{3/2+b_0\alfs}(1-2x)+
b_2A_2C_4^{3/2+b_0\alfs}(1-2x)\right)
\rlabel{sumrdf}
\een
in the NNA approximation, and \rref{sumrdf} with $\alpha _s =0$ in the
one loop approximation. In \rref{sumrdf}, $A_i({\alpha _s}) $ are given
by \rref{An}.
In the one loop
expansion, \rref{1lsol},this implies $b_2 (\mu _0 ^2) = 1.0$, and
in the NNA expansion, \rref{NNAdf} $b_2 (\mu _0 ^2) = 1.6$. We have
represented graphically both functions in fig.\ref{distf}.
We agree with \rcite{BF} that the higher
harmonics remove the oscillations and make the function smoother.

\begin{figure}[h]
\begin{center}
\leavevmode\epsfxsize=13cm\epsfbox{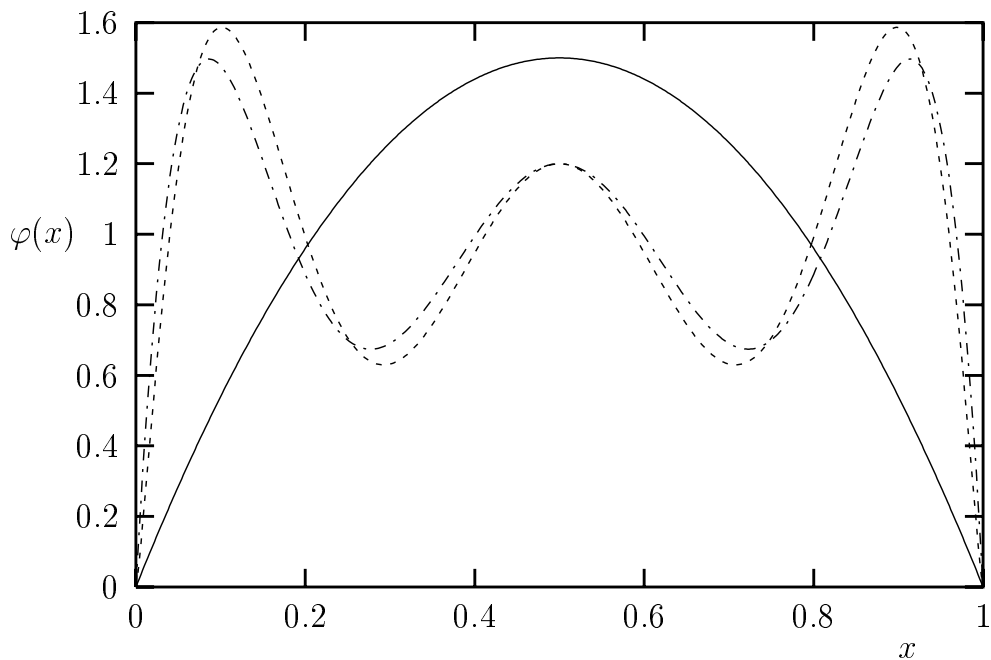}
\end{center}
            \caption{ Several distribution functions.
Solid line: asymptotic distribution function.
Dot-dashed line: one-loop distribution function obtained by combining
sum rule results \rref{mom1} and the prediction \rref{1/2}.
Dashed line: NNA distribution function, (\ref{sumrdf}) obtained
as above. Here,
$\mu _0 ^2 =$ 1 GeV$^2$.
\rlabel{distf} }
\end{figure}

The evolution of the moments $b_k (\mu ^2)$ with $\mu ^2$
is given by \rref{mixed}.
It can be shown the in the limit $\mu \rightarrow \infty$, all
$b_{i>0}(\mu^2)$ tend to zero. This follows from the fact
 that at  the one loop level the NNA approximation reduces to the exact
one loop approximation. Then only the lowest
momentum, $b_0(\mu^2)$ will be relevant in the limit 
$\mu \rightarrow \infty$. This leads to 
the asymptotic wave function \footnote{Another way of showing that 
$b_{i>0}(\mu^2)$ tends to zero in the limit $\mu^2 \rightarrow \infty$ 
is by induction on $i$: First, we solve \rref{b1ECUA}, and show that 
$b_1(\mu^2)$ x vanishes in the limit $\mu^2 \rightarrow \infty$. Then, 
the equation for $b_2$ can be solved, implying that $b_2$ also tends to 
zero. This is then repeated for $b_3,\ b_4, \cdots$}:

\bee
\varphi_{as}(x)=\lim_{\mu \rightarrow \infty}
 \frac{\Gamma(4+2b_0\alfs)}{\Gamma^2(2+b_0\alfs)}
[x(1-x)]^{1+b_0\alfs}=6x(1-x). %\ \alfs\equiv\alfs(\mu^2) .
\rlabel{lmphi}
\een
It is interesting to see the difference between the asymptotic shape
and the lowest order NNA harmonic, which depends on $\alpha _s$.
In fig.\ref{asmphi} we can see that the lowest NNA harmonic
is a little bit narrower than $\varphi _{as} (x)$.

\begin{figure}
\begin{center}
\leavevmode\epsfxsize=13cm\epsfbox{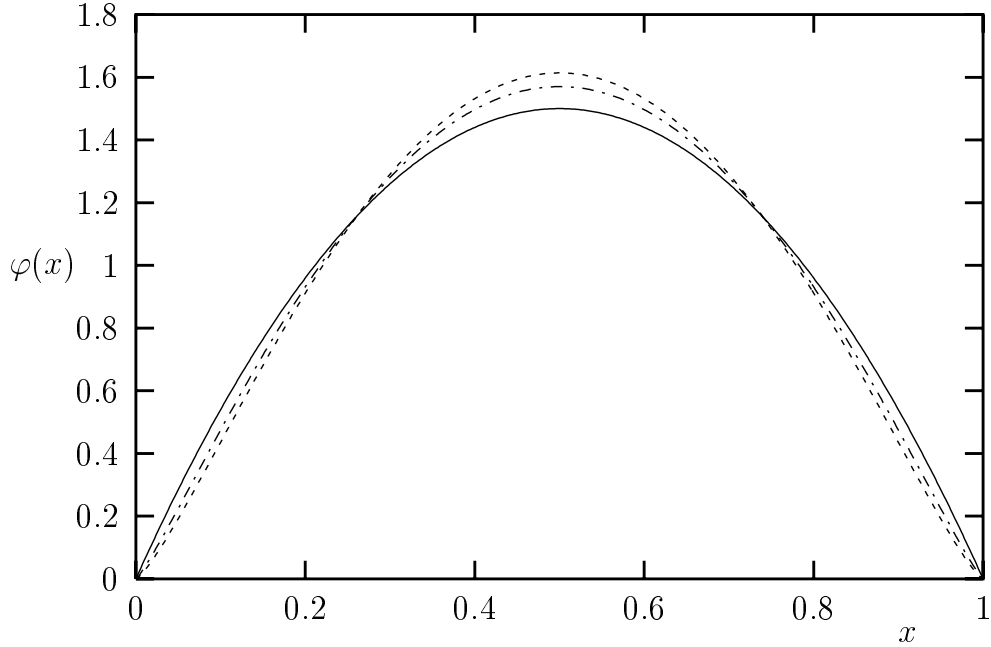}
\end{center}
            \caption{
  Solid line: asymptotic distribution function.
  Dashed line: Asymptotic NNA distribution function
(\ref{lmphi}) for $Q^2=10 {\rm GeV}^2$.
Dot-dashed line:  the same for $Q^2=1  {\rm GeV}^2$.
\rlabel{asmphi} }
\end{figure}

The effect of the nondiagonal terms $C_{k \ l}$ on the evolution of
the coefficients of the two models presented in this subsection
has been displayed in
fig.\rref{mix}.c and fig.\rref{mix}.d. In fig.\rref{mix}.c we have
solved the differential equation \rref{mixed} exactly for the initial
condition corresponding to \rref{lmphi}, that is,
$b_0(1$ GeV$^2)$ = 1, $b_{i>0}(1$ GeV$^2)$ = 0. Here, the
diagonal solution \rref{bUk} predicts
$b_{1}(\mu^2) = b_{2}(\mu^2)=0$, while the exact solution is negative,
and quite small. In fig.\rref{mix}.d we have used the initial conditions
the nondiagonal elements $C_{k \ l}$ for the initial condition
$b_0(1$ GeV$^2)$ = 1,  $b_1(1$ GeV$^2)$ = 1.7 and
$b_2(1$ GeV$^2)$ = 1.6, which is obtained by combining a sum rule
prediction for the first moment, \rcite{BF,CZ}, and a prediction for
$\varphi (1/2)$, \rcite{BF}.
The moments $b_k (\mu)  ^2$ are the matrix elements of the
operators
\bee
  \bar d\gamma_5(\hat{n})(n\partial)^{2n}_{+}
C^{3/2+b_0\alfs}_{2n}(nD_{-}/(n\partial)_{+})u.
\rlabel{NNAgo}
\een
which renormalize multiplicatively.
It might be convenient to redefine the one loop operators
 (\ref{1lgo}) in order to extract from the diagonal part
all the main corrections.

\subsection{Conformal symmetry constraints}

We will now make some comments on the diagonalization of the evolution
kernel and its eigenfunctions based on conformal invariance arguments.
We start with the observation that at the one loop level we have
conformal invariance, which gets broken when we go to higher orders.
Conformal symmetry is broken by the introduction of the renormalization
%%%%%%%%%%%%%%%%%%%%%%%%%%%%%%%%%%%%%%%%%%%%%%%%%%%%%%%%%%%%%
% CHANGE BEGINS
%%%%%%%%%%%%%%%%%%%%%%%%%%%%%%%%%%%%%%%%%%%%%%%%%%%%%%%%%%%%%
scale $\mu$, and due to renormalization scheme effects \rcite{Muller1}.
The breaking due
to the renormalization of the coupling are, to our accuracy,
proportional to the first
coefficient of the $\beta -$ function, while those due renormalization
scheme 
effects are $1/N_f^2$ suppressed, and therefore lie beyond our accuracy.
%%%%%%%%%%%%%%%%%%%%%%%%%%%%%%%%%%%%%%%%%%%%%%%%%%%%%%%%%%%%%
% CHANGE ENDS
%%%%%%%%%%%%%%%%%%%%%%%%%%%%%%%%%%%%%%%%%%%%%%%%%%%%%%%%%%%%%
Despite the fact that conformal symmetry is broken, it still allows us
to diagonalize the evolution kernel.  This is equivalent to the
diagonalization of the anomalous dimension matrix.
The fact that the anomalous dimension matrix can be diagonalized at
any order
in perturbation theory follows from conformal invariance at the one
loop level, where \rref{1lad} is implied. In general, the eigenvalues
will be given by a series in $\alpha _s$. Since there are
no two eigenvalues with the same lowest order coefficient, \rref{1lad},
all eigenvalues are different, and we can therefore diagonalize the
anomalous dimension matrix.

Consider now the renormalization group equations for the operators
defined in (\ref{1lgo}) for higher orders in perturbation theory:
\bee
\mu^2\frac{d}{d\mu^2}O_R=\hat\gamma O_R ,
\rlabel{RG}
\een
where $O_R$ are the renormalized operators, and $\hat\gamma$ is the
anomalous dimension matrix, which is only diagonal at the one loop
level. Since it can be diagonalized, there is a matrix $\hat U(\alfs)$
such that
\bee
\hat\gamma(\alfs)=\hat U^{-1}(\alfs)\hat\gamma_D(\alfs)\hat U(\alfs)
\rlabel{diaGAMMA}
\een
where $\hat\gamma_D$ is a diagonal matrix.
Substituting this representation in \rref{RG} and redefining the
operators
\bee
\tilde O_R(\alfs) = \hat U (\alfs)O_R
\een
it is easy to obtain the following equation for the
new operators $\tilde O_R$:
\bee
\mu^2\frac{d}{d\mu^2}\tilde O_R=\left\{
\hat\gamma_D(\alfs)-
\hat U (\alfs)\left(\mu^2\frac{d}{d\mu^2}\hat U^{-1}(\alfs)\right)
\right\} \tilde O_R
\rlabel{RGtilde}
\een
In the rhs, the coefficient in front of $\tilde O_R$ is its anomalous
dimension. It is clear that to our accuracy, we can associate the
diagonalization of the matrix $\hat\gamma$
with the diagonalization of the kernel $V_\alpha(x,y)$. Then, the
nondiagonal part, $\hat U (\alfs)\left(\mu^2\frac{d}
{d\mu^2}\hat U (\alfs)\right)$  is the mixing matrix $\hat C(\alfs)$
and the operator $\tilde O_R(\alfs)$ is given in (\ref{NNAgo}).

In $D=4-2\epsilon$ dimensions, the $\beta$-function has
the following form:
\bee
\beta(\epsilon,\alfs)= -\epsilon \alfs - b_0\alfs^2 + \cdots
\rlabel{beta}
\een
and there is a critical value for the coupling (or fixed point )
 $\alfs^\ast$ such that
\bee
\beta(\epsilon,\alfs^\ast)= 0
\rlabel{zero}
\een
To first order in the large $N_f$ expansion, (\ref{beta}) implies
\bee
\alfs^\ast = \frac{3\epsilon}{2N_f}(4\pi) + O(1/N_f^2)
\een
Consider now equation (\ref{RGtilde}) at the critical point.
Than nondiagonal part  vanishes:
\bee
\hat U (\alfs)\left(\mu^2\frac{d}{d\mu^2}\hat U^{-1} (\alfs)\right)
|_{\alfs=\alfs^\ast}=
\beta(\epsilon,\alfs)\hat U
 (\alfs)\left(\frac{d}{d\alfs}\hat U^{-1}(\alfs)\right)
|_{\alfs=\alfs^\ast}=0,
\een
due to (\ref{zero}), and $\tilde O_R(\alfs^\ast)$ has a
diagonal anomalous
dimension matrix $\hat\gamma_D(\alfs^\ast)$ at the critical point.
For the operators
(\ref{NNAgo}),
taking into account that we have to perform the substitution
$2/3N_f \rightarrow 2/3N_f - 11/3N_c$ ``by hand'',
and then
$\alfs^\ast =-\epsilon/b_0$, we obtain
 \bee
\tilde O_R(\alfs^\ast)=
  \bar d\gamma_5(\hat{n})(n\partial)^{2n}_{+}
C^{3/2+b_0\alfs^\ast}_{2n}(nD_{-}/(n\partial)_{+})u=
  \bar d\gamma_5(\hat{n})(n\partial)^{2n}_{+}
C^{3/2-\epsilon}_{2n}(nD_{-}/(n\partial)_{+})u
\rlabel{NNAcgo}
\een
The last equation shows that the basis (\ref{NNAgo}) is conformal
at the critical point because
the classical
conformal composite operator, built of two fermion fields
$\bar\psi,\ \psi$ is given by \rcite{conf}:
\bee
  \bar \psi \gamma_5(\hat{n})(n\partial)^{2n}_{+}
C^{d_\psi}_{n}(nD_{-}/(n\partial)_{+})\psi \ ,
\een
where $d_\psi$ is the canonical dimension of the fermion field.
In $D=4-2\epsilon$ dimensions, $d_\psi=3/2-\epsilon$.
We have checked by explicit calculation that at the critical
point, $\alfs^\ast = \frac{3\epsilon}{2N_f}(4\pi)$,
and to leading order in
the large $N_f$ expansion, the operators
\rref{NNAcgo} have a diagonal anomalous dimension matrix
$\gamma_D(\alfs^\ast)$\footnote{Recently this anomalous dimension at the
critical point has been calculated in \rcite{Gracey}}.

We conclude that conformal symmetry operators at the critical point
and the operators $\tilde O_R(\alfs)$ that diagonalize the evolution
kernel at leading order in $1/N_f$ are related:
The operators $\tilde O_R(\alfs)$ can be obtained from the
the conformal ones by the replacement
$( -\epsilon ) \rightarrow b_0\alfs$. We should notice that this
result is rigorous at leading order in the $1/N_f$ expansion.
However, we follow the prescription of NNA, and replace
$2/3N_f \rightarrow 2/3N_f - 11/3 N_c$.
It seems reasonable to assume that the nondiagonal part of the
anomalous dimension matrix induced by this will be small.

\section{The form factor }

Below we calculate the form factor \rref{fpi} for  $\omega=1$.
In this case, \rref{fpi} reads:
\bee
   F_{\pi\gamma}(Q^2)\ =\frac{N}{Q^2}\int^1_0
   \left(C_0(x) + C_1(x,Q^2)+C_0(1-x) + C_1(1-x,Q^2) \right)
   \varphi(x,Q^2)dx\ , \
 \rlabel{NNAfpi}
\een
where for simplicity we set $\mu^2=Q^2$.
 $C_0(x)$,\  $C_1(x,Q^2)$ can be obtained from
\rref{Borncf} and \rref{NNAcf}, and $\varphi(x,Q^2)$
is given by \rref{NNAdf}.

We have already seen that the c\cf {} \rref{NNAcf} is ill defined
in the limit $x\rightarrow 0$ and $x\rightarrow 1$ because we
expand in $\Lambda^2/(xQ^2)$ or
$\Lambda^2/((1-x)Q^2)$. This induces uncertainties which are
related to the nonperturbative structure of QCD
in the infrared region and induce power corrections
to the form factor.
In addition to the power contributions from the $IR$-renormalon
ambiguity, the form factor
can get qualitatively different power corrections from the regions
$x\rightarrow 0$ and $x\rightarrow 1$. These new power corrections can
be related to new contributions which correspond to
the new regimes, see fig.\ref{regimes}.2 and \ref{regimes}.3.
 As discussed above, the
leading contributions of these regimes are of order $1/Q^4$ and
become essential in the limit $\omega\rightarrow 1$. To perform a
phenomenological analysis of the power uncertainty related to
the regions $x\rightarrow 0$ and $x\rightarrow 1$, we first
observe that the all the ambiguities come from the Borel
integral of $C_1(x,Q^2)$, see \rref{NNAcf}, and interchange the
order of integration, that is, integrate first over $x$, and
then over the Borel parameter $u$.  Using the property
$\varphi(x)=\varphi(1-x)$ we rewrite \rref{NNAfpi} as:
 \bee
 F_{\pi\gamma}(Q^2)\ =2\frac{N}{Q^2}\int^1_0
  \left( C_0(x) + C_1(x,Q^2)  \right)\varphi(x,Q^2)dx\ .
 \een
   The complete integration over $x$ can be
 done analytically. In principle, it is only necessary to interchange
the order of integration in the Borel integral in
\rref{NNAcf}, because the other integral is well defined.
     We obtained an expression for the form factor in which
     all ambiguities are  related to the poles of the
     function in the Borel integral. The integration over $x$ yields
     (below $\alpha \equiv 1+b_0\alfs$):
 \begin{eqnarray}
 &&F_{\pi\gamma}(Q^2)\ =\
{N \over Q^2} \sum^\infty_{k=0}\frac{(2k)!}{(2\alpha)_{2k}}
\frac{(1+2\alpha+4k)}{(\alpha+k)}b_{k}(Q^2)
\left\{1-{C_F \over 4 \pi b_0} M_{k}(\alfs) \right\}\ ,
\rlabel{NNAff} \\
&&\ M_{k}(\alfs)\ =\ p.v.\int\limits^\infty_0 e^{-u/b_0\alfs}
\frac{du}u \left\{e^{u(C-ln2)} \frac{2 \Gamma(\alpha-u)}
{(1-u)(2-u)\Gamma(\alpha)}\theta_k(\alfs,u)
-\theta_k(\alfs,0)\right\} \rlabel{Mk}  \nonumber \\
&&  + \  \
\int^\alpha_1\ \frac{d\lambda}{1-\lambda}
\{\rho_k(\alfs,\lambda)-\rho_k(\alfs,1)\}\ ,   \\ \nonumber \\
&&\theta_k(\alfs,u) = 3\frac{(1+u)_{2k}}{(2k)!}\
\frac{\Gamma(1+2\alpha+2k)}{\Gamma(1+2\alpha-u+2k)}+
\frac{(1+u)_{2k}}{(2k)!} \frac{(2-u)(1+u)(\alpha-u)}{(1+u+2k)}
\nonumber \\
&& \times \frac{\Gamma(1+2\alpha+2k)}{\Gamma(2+2\alpha-u+2k)}
{}_3 F_2 \left[ \begin{array}{l}
1,1,\ 1+\alpha+2k\\ 2+u+2k, 2+2\alpha-u+2k \end{array} |1\right]
\nonumber \\
&& -\ \ \frac{2\Gamma(1+2\alpha+2k)}{\Gamma(1+2\alpha-u+2k)}
\sum^{2k}_{\ell=0}\
\frac{(1+u)_\ell}{(1+u+\ell)\ell!}\ _3 F_2 \ \left[
\begin{array}{l} 1,1,\ 1+\alpha+2k \\ 2+u+\ell, 1+2\alpha-u+2k
\end{array}| 1\right]\ . \nonumber
 \end{eqnarray}
\begin{eqnarray}
\rho_k(\alfs,\lambda)&=&\frac{\Gamma(2+2\lambda)}{3\Gamma(2-\lambda)
\Gamma^3(1+\lambda)}\left\{\frac{2+2\lambda-\lambda^2}{1+\lambda} +
\right. \nonumber \\ &+& \frac{\alpha\lambda^2}{(\lambda+2k)(1+2\alpha+2k)}
\ _3F_2 \left[
\begin{array}{l} 1,\lambda,\ 1+\alpha+2k \\ 1+\lambda+2k, 2+2\alpha+2k
\end{array}|1 \right] \nonumber \\
 &-& \left. \sum^{2k}_{\ell=0}\ \frac\lambda{\lambda+\ell}\ _3F_2
\left[\begin{array}{l} 1,\lambda,\ 1+\alpha+2k \\  1+\lambda+\ell,
1+2\alpha+2k \end{array} |1\right] \right\}\ .
\end{eqnarray}
Where $_3F_2[\cdots|1]$ is the standard hypergeometric function (3,2),
$(z)_{n}=\frac{\Gamma(z+n)}{\Gamma(z)}$ is the Pochhammer symbol,
and $C_F=(N_c^2-1)/(2N_c)$. We have absorbed all the singularities
in the factor
${\Gamma(1+b_0\alfs-u)} / [{\Gamma(1+b_0\alfs)(1-u)(2-u)}]$ of
the Borel integral in \rref{Mk}.  There are two renormalon poles at
$u=1$ and $u=2$, and an infinite number of new ``small-$x$'' poles at
$u=1+b_0\alfs+m, \ m=0,1,\cdots$. We will integrate the poles following
the principal value prescription. The final result will be prescription
dependent, but this dependence is known to be power suppressed.

  To estimate the size
 of the higher power corrections, we will follow the recipe of
 Section 2: A pole at $u=u_0$ induces a power correction of order
 $( \Lambda ^2 / Q ^2 )^{u_0}$. The entire power
 correction to the form factor can written in the following way:
 \beqn
 \Delta F_{\pi\gamma}(Q^2) &=&
 \pm \frac{N}{Q^2}
 \left( N_1 \Delta_2(\alfs) \left(\frac{\Lambda}{Q}\right)^2 +
        K_1 \delta_2(\alfs)
        \left(\frac{\Lambda^2}{Q^2}\right)^{1+b_0\alfs} + \cdots
        \right) =  \nonumber  \\ &\pm &\frac{N}{Q^2} \left(
 \{N_2 \Delta_2(\alfs) + K_2 \delta_2(\alfs) e^{-1}\}
 \left(\frac{\Lambda}{Q}\right)^2
        + \cdots \right) \ .
\rlabel{powcor}
 \eeqn
In the last identity we have used the
one-loop expression for the running coupling.
By the dots, we denote the contributions of the higher poles,
$u=2,\ \mbox{and} \ u=1+b_0\alfs+m, \ m=1,2,\cdots$.
The contribution $N_1\Delta_2(\alfs)$ is
related to the first renormalon pole, $u=1$.
The function $\Delta_2(\alfs)$ is build of the residues of the
integrand of the
Borel integral, and can be calculated, and $N_1$ is an unknown number
of order one. The second contribution is related to the
``small-$x$'' poles.
Its structure  is the same, $K_1$ is an unknown constant
and $\delta_1(\alfs)$ a calculable function.
We finally obtain:
\beqn
 F_{\pi\gamma}(Q^2)\ &=&\frac{N}{Q^2} \left(
\sum^\infty_{k=0}\frac{(2k)!}{(2\alpha)_{2k}}
\frac{(1+2\alpha+4k)}{(\alpha+k)} b_{k}(Q^2)
\left\{ 1-C_F/b_0M_{k}(\alpha) \right\} \right. \nonumber \\
&\pm &
 \left. \{ N_2 \Delta_2(\alfs) + K_2 \delta_2(\alfs) e^{-1}\}
 \left(
 \frac{\Lambda}{Q}
 \right)^2 + \cdots
  \right).
 \rlabel{ff}
\eeqn
Here, we are neglecting  $1/Q^4$ and higher power contributions.
To make a numerical estimation, we only take, for simplicity, the first
term of the  expansion of the d\df {} \rref{lmphi}
\bee
 \varphi(x,Q^2)=\frac{\Gamma(4+2b_0\alfs)}{\Gamma^2(2+b_0\alfs)}
[x(1-x)]^{1+b_0\alfs}, \ \alfs\equiv\alfs(Q^2) .
\rlabel{numphi}
\een
As discussed above, this approximation is very close to the
asymptotic form. Then, in the
expression for the form factor \rref{NNAff}, we have to keep only  the
first term. This gives:
\begin{eqnarray}
&&F_{\pi\gamma}(Q^2)\ = \frac{N}{Q^2}
\frac{(1+2\alpha)}{ \alpha}\left\{1-{C_F\over 4 \pi b_0}
M_{0}(\alfs)
\right\} \pm \mbox{power corrections}
\rlabel{numff} \\
&&\ M_{0}(\alfs)\ =\ p.v.\int\limits^\infty_0
e^{-u/b_0\alfs} \frac{du}u \left\{e^{u(C-ln2)}
\frac{\Gamma( \alpha-u)}{\Gamma(\alpha)}
\frac{2\theta_0(\alfs,u)}{(1-u)(2-u)}
-\theta_0(\alfs,0)\right\} \nonumber  \\ && +\ \
\int^\alpha_1\ \frac{d\lambda}{1-\lambda} \{\rho_0 (
\alfs,\lambda)-\rho_0(\alfs,1)\}\ , \\ \\
&&
\theta_0(\alfs,u) =\Gamma(1+2\alpha) \left\{
\frac{3}{\Gamma(1+2\alpha-u)}+
\frac{(2-u)(\alpha-u)}{\Gamma(2+2\alpha-u)} {}_3 F_2 \left[ \begin{array}{l}
1,1,\ 1+\alpha \\ 2+u, 2+2\alpha-u \end{array}|1 \right]
\right.
\nonumber \\
&&\left. -\ \ \frac2{\Gamma(1+2\alpha-u)(1+u)}{}_3F_2 \ \left[
\begin{array}{l} 1,1,\ 1+\alpha \\ 2+u, 1+2\alpha-u
\end{array}|1\right]\ \right\}
\end{eqnarray}

\begin{eqnarray*}
\rho_0(\alfs,\lambda)&=&\frac{\Gamma(2+2\lambda)}{3\Gamma(2-\lambda)
\Gamma^3(1+\lambda)}\left\{\frac{2+2\lambda-\lambda^2}{1+\lambda} +
\frac{\alpha\lambda}{(1+2\alpha)} \right. \\
&\times&_3F_2 \left[
\begin{array}{l} 1,\lambda,\ 1+\alpha \\ 1+\lambda, 2+2\alpha
\end{array} | 1 \right]
 \left.-{}_3F_2
\left[\begin{array}{l} 1,\lambda,\ 1+\alpha\\ 1+\lambda,
1+2\alpha \end{array} | 1 \right] \right\}\ .
\end{eqnarray*}
At leading order, we obtain
\bee
F_{\pi\gamma}^{LO}(Q^2)\ = 3\frac{N}{Q^2}
\rlabel{ldo}
\een
To obtain the next-to-leading, or 1-loop result, we expand
\rref{numphi} in series of $\alfs$:
\beqn
\varphi(x,Q^2)=6x(1-x)\{1+b_0\alfs(Q^2)[5/3+\ln(x(1-x))]\}+O(\alfs^2).
\eeqn
The coefficient function \rref{NNAcf} in the one loop limit
gives the  well known one-loop formula \rref{1lcf}, and
to next-to-leading order in $\alfs$, we obtain:
\beqn
F_{\pi\gamma}^{NLO}(Q^2)\ &= &\frac{N}{Q^2}\left\{ 3 +
6 \alfs b_0 \int_0^1 (1-x)\bigg[{5\over 3}+\ln[x(1-x)] \bigg] dx \right.
 \nonumber \\
&+& \left.
C_F\frac{\alfs}{4\pi}\int_0^1 (1-x)
\bigg[ \ln2(3+2\ln x)+\ln^2(x)-\ln(1-x)-9 \bigg] dx \right\} =
\nonumber \\
&=& 3 \frac{N}{Q^2} \left( 1-\frac13b_0\alfs-
5 C_F {\alfs \over 4 \pi} \right)
 \rlabel{NLOff}
\eeqn
The contribution of higher order radiative corrections
in the NNA approximation
is given by \rref{numff}.
Numerical results are presented in fig.\ref{number}. For
$\alfs(Q^2)$, we use $\alfs(m_{\tau}^2)=0.32,\ N_f=4$, and take
into account that in the $\overline{MS}$-scheme $C=5/3$.  We see
that the one loop correction decreases the leading order result
by $20\%$. Higher order contributions, together with the
one-loop correction, decrease the leading order by $30\%$. This
estimation is in agreement with the estimations for radiative
corrections suggested in \rcite{MURA}.

\begin{figure}[t]
\begin{center}
\leavevmode\epsfxsize=13cm\epsfbox{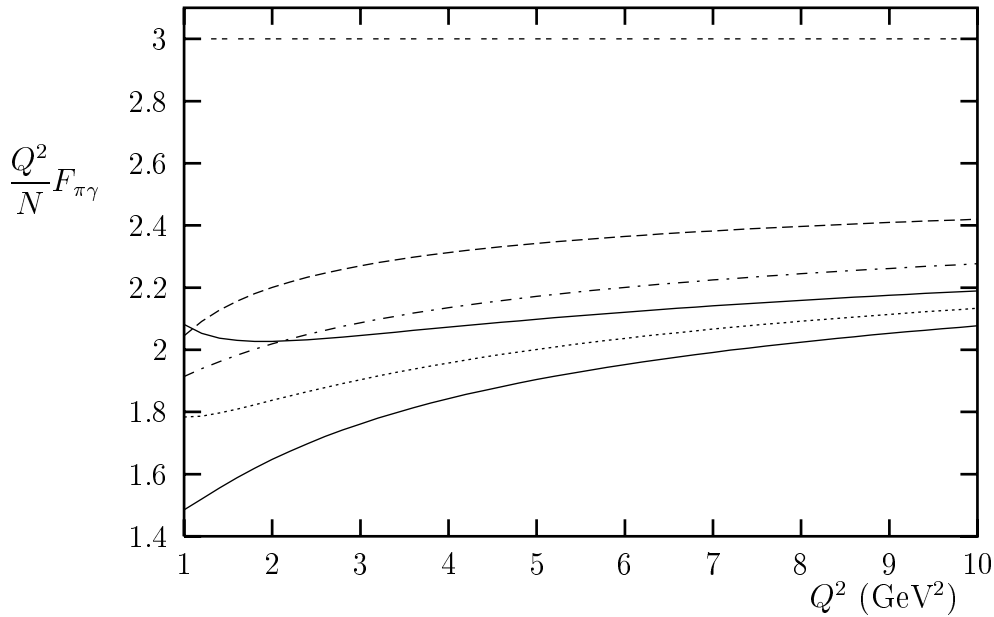}
\end{center}
            \caption{
The form factor $F_{\pi \gamma }$.
The constant line $(Q^2/N)F_{\pi \gamma }=3$ corresponds to the lowest
order prediction, (\ref{ldo}). The next to leading result, (\ref{NLOff})
is plotted with long dashes, while the dots represent the NNA prediction,
(\ref{numff}), and the two solid lines represent the uncertainties in
the resummation. The dot-dashed line is the semisum of the NNA result
and the next to leading prediction, and we expect the real result not
to lie far away from it.
 \rlabel{number} }
\end{figure}

 The difference between our d\df {} and
 the asymptotic one is very small and,
for example, a discrimination in the description of the NLO correction
to the form factor will be
very difficult. Let us remind that  in \rcite{MURA} it has
been assumed that the corrections due to the
evolution of the d\df {} are so small that they
can be neglected.

To obtain the power corrections, we set the unknown
 constants $K_2$ and $N_2$ in
\rref{powcor} equal to 1. We find that in our approach the
power corrections
give an ambiguity of order $10-2\%$ for
$2 {\rm GeV}^2<Q^2<10{\rm GeV}^2$.
Below we give some numerical results in the following form:
\bee
N^{-1}Q^2F_{\pi\gamma}(Q^2)=
F_{\pi\gamma}^{NLO}(Q^2)-(\mbox{number1})\pm(\mbox{number2}),
\een
where ``number1'' is the contribution of the higher order radiative
 corrections in the
NNA approximation, and ``number2'' is the power correction
(ambiguity in the summation of the
perturbation series).

\begin{table}[h]
\begin{center}
\begin{tabular}{|c|c|} \hline
$Q^2$ $(\rm{GeV})^2$ & $N^{-1}Q^2F_{\pi\gamma}$ \\ \hline
2  & {2.20}-{0.36}$\pm${0.19}  \\ \hline
3  & {2.27}-{0.36}$\pm${0.14}  \\ \hline
4  & {2.31}-{0.35}$\pm${0.12}  \\ \hline
5  & {2.34}-{0.34}$\pm${0.10}  \\ \hline
6  & {2.36}-{0.33}$\pm${0.08}  \\ \hline
7  & {2.38}-{0.32}$\pm${0.07}  \\ \hline
8  & {2.40}-{0.31}$\pm${0.07}  \\ \hline
9  & {2.41}-{0.30}$\pm${0.06}  \\ \hline
10 & {2.42}-{0.29}$\pm${0.06}  \\ \hline
\end{tabular}
\end{center}
\caption{Some numerical results for the form factor}
\rlabel{fpiTABLE}
\end{table}
For the last number, see \rref{powcor},  we set
$N_2=N_4=K_2=K_4=1$
(in this calculation we take into account small
 terms of order $(\Lambda/Q)^4$).  Such choice fixes the
relative sign of renormalon and small x ambiguities which is
also unknown.  In Fig.\ref{sepPOLES} we have plotted the separate
contribution of the first 4 poles to the ambiguity. It is
interesting to notice that the first renormalon pole ambiguity,
due to the $u=1$ singularity, and the first small $x$ ambiguity,
due to the $u=1+\alfs b_0$ have different sign, which results in
a partial cancelation of the ambiguity. This cancelation
suggests  that the contributions of the matrix elements of the
higher twist operators will partially cancel with contributions
coming from photon emission at large distances.

Other choice, such as for example $N_2=-K_2=1$,
which would lead to a bigger uncertainty ($40\%$ at $Q^2=2$ GeV$^2$
and $10\%$ at $Q^2=10$ GeV$^2$ ),  could also have been adopted.
At present we have no rigorous arguments to choose the relative
sign. Any of these definitions can only provide us with an
estimation for the uncertainty.

\begin{figure}
\begin{center}
\leavevmode\epsfxsize=12cm\epsfbox{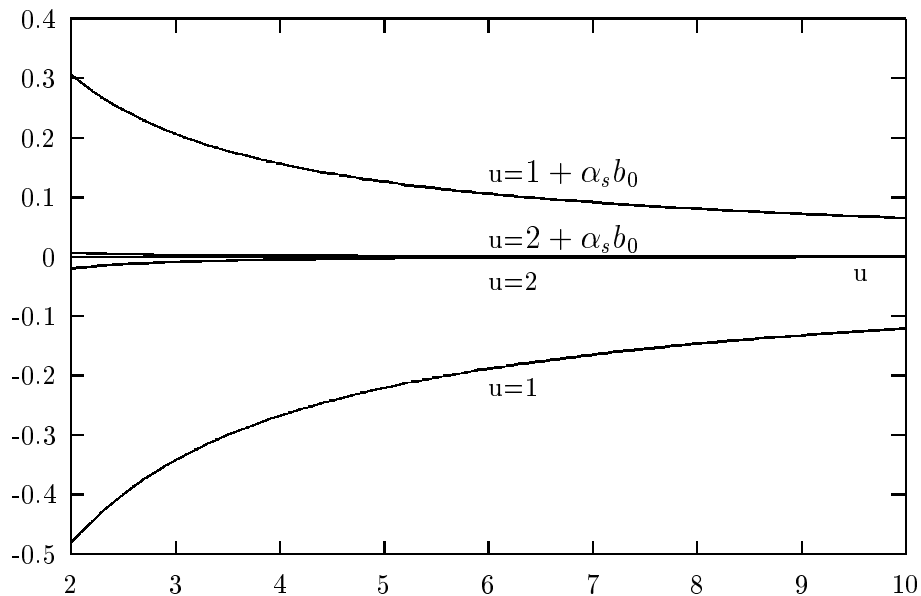}
\end{center}
\caption{The separate contribution of each pole to the uncertainty.
\rlabel{sepPOLES}}
\end{figure}

As a rule, NNA resummations give an exceeding estimation for
the higher order
 contributions, see for example \rcite{BBB}, and the real curve
 for the form factor might lie a little above the one suggested
 by the NNA approximation in fig.\ref{number} but below the one
 loop approximation. We have plotted an estimation for the real
 curve in the fig.\ref{number}  with long dashed (short gaps)
 line.

 To understand better the shape of the distribution function, we have
 to consider at least the model of the d\df {} \rref{sumrdf} which was
 obtained from the consideration of the sum rule constraints. This is
work in progress, \rcite{BGK}.
%But this is the subject of another paper \rcite{BGK}.

\section{Conclusions}
We have obtained the photon $\pi ^0$-meson form factor at leading order
in `naive nonabelianization' (NNA).
We have computed the coefficient function and the evolution kernel that
governs
the evolution of the distribution function, in the NNA approximation.

 To evaluate the c\cf{} and the evolution kernel, all leading
contributions in $(-b_0 \alpha _s)^n$ have to be computed and resummed.
For the c\cf, Borel integral techniques have been used to perform
the summation. Due to the presence of (IR) renormalons in the integrand,
a prescription to perform the integration has to be chosen. We have
used the principal value prescription.
The assumption of ultraviolet dominance of higher twist matrix element
has  allowed us to obtain an estimation of higher twist contributions
(ambiguity in the resummation).
We have found that when one of the photons is on shell,
the c\cf {} has new singularities.
These new singularities are caused
by new IR regions, that appear when one of the quarks that build
the pion carries almost no momentum. The power correction induced by
these new singularities have been taken into account.

We have found that conformal constraints allow to find a basis of
operators that diagonalizes the evolution kernel in the NNA
approximation, but the anomalous dimension matrix is nondiagonal.
The nondiagonal terms of the anomalous dimension matrix have been found
to be much smaller than the diagonal elements and we neglected by them.
We have obtained an expansion of the d\df {} in series of
Gegenbauer polynomials with multiplicatively renormalized moments. This
form is very convenient to analyze the effects of radiative corrections.
Radiative corrections make the shape of the first term in the expansion
of the d\df{} a little bit
narrower as compared to the one loop term but, we
find that  this effect is small.
We have calculated the convolution of the c\cf {} and the d\df{},  and
obtained
the form factor as a sum with unknown nonperturbative
coefficients, which depend on the shape of the wave function in the
low energy region.
In our numerical analysis, we only keep the first term in the
expansion of the d\df.
We obtained that the radiative correction decrease the
leading order by $30\%$. This is consistent with the estimation for the
radiative correction by \rcite{MURA}.
The value of the power suppressed corrections (ambiguity in the
resummation) depends on the relative sign of renormalon and
small x ambiguities. It will be smaller  for the
case when they have different sign and predict the uncertainty
for the value of the form factor from $10\%$ at $Q^2=2{\rm
GeV}^2$ to $2\%$ at $Q^2=10{\rm GeV}^2$. The different sign of
the ambiguities implies that nonperturbative effects coming from
the matrix elements of the higher twist operator and soft photon
emission is partially cancelled. In case of the similar sign
the uncertainty for the value of the form factor is bigger:
 from $40\%$ at $Q^2=2{\rm GeV}^2$ to $10\%$ at $Q^2=10{\rm
 GeV}^2$.
Clearly, an investigation of the influence of the
higher harmonics in the expansion of the d\df{} is needed. This
will allow us to understand better the structure of the d\df{}
in the low energy region.

{\bf Acknowledgments.} We are grateful to V.~Braun., who initiated
this work, for the numerous discussions and critical reading of this
paper. We also thank S.~Peris for the critical reading of the manuscript,
and A. Radyushkin and D. M\"uller for discussions. 
 N.~K. thanks NORDITA for hospitality where the part of this
work has been done. P.~G. acknowledges gratefully a grant from the
Spanish ministry for education and culture (MEC).

\appendix

\renewcommand{\theequation}{\Alph{section}.\arabic{equation}}

\section*{Appendix A}
\setcounter{equation}{0}
\addtocounter{section}{1}

All calculations have been performed in dimensional regularization in
$D=4-2\epsilon$ dimensions. For the gluon propagator, we use the Landau
gauge.
To solve the $\gamma_5$ - problem, we have followed the prescription
of \rcite{braaten}, where an anticommuting $\gamma _5$ was only used in
the box diagram.
The unrenormalized contribution of each diagram $D_i$  to the c\cf {}
can be represented in the following way:
\beqn
D_i  &=& -\frac{C_F}{4 \pi}
 \sum ^{\infty}_{n=0}\frac{1}{(1+n)\epsilon^{(n+1)}}\alfs^{1+n}(\mu^2)
(-\beta_0^f)^n E_i[(n+1)\epsilon,\epsilon]\ ,
\rlabel{allblobs}\\ \nonumber
E_i[u, \epsilon] &=& \left(4\pi \frac{\mu^2}{Q^2}\right)^u
\left( \frac{\Gamma (4-2\epsilon)}
{6\Gamma (1+\epsilon)\Gamma^2(2-\epsilon)}  \right)^{1-\frac{u}{\epsilon}}
\frac{\Gamma (1-\epsilon)\Gamma (1+u)\Gamma (1-u)}
{\Gamma (2-\epsilon+u)\Gamma (2-\epsilon-u)}
R_i(\epsilon,u|x,\omega) \ , \nonumber
\eeqn
where  $n$ is the number of fermion bubbles in the gluon propagator,
 $Q^2=-(q_1-q_2)^2/4>0,\ \beta_0^f\ =\ \frac23N_f/4\pi$ is
 the fermion part of the full $\beta_0=-b_0$,
 and only the factor $R_i(\epsilon,\Delta,x,\omega)$
is different for different diagrams.
We will refer to the contribution of the $g_{\mu \nu}$ term in the
gluon propagator as the ``gauge invariant'' contribution, and to the
 contribution from the  $k_\mu k_\nu / k^2$ part as
the ``gauge dependent'' contribution. The total contribution of the
``gauge dependent'' cancels, and we will in what follows only deal
with the ``gauge invariant'' ones. In what follows, the notation
\be
\alpha = 1-\epsilon,\quad
 F(z)\equiv {}_2F_1(1,\alpha; 1+\alpha+u|z)
\ee
where ${}_2F_1$ is the  hypergeometric function (2,1),
will be understood.

The contribution of the crossed diagram of $R$ will be denoted by
 $\bar{R}$,  $C_0(x,\omega)$ denotes the Born contribution, see
(\ref{Borncfw})). The diagrams are numbered according to
fig.\ref{1ldiagramms}:
 \beqn
 R_1+\bar{R}_1 &=& C_0^{1+u}(x,\omega) \frac{2\alpha^2(\alpha+u)}{1+\alpha-u}
  +\{ x\leftrightarrow 1-x\} \ , \\ \nonumber
 R_2+\bar{R}_3 &=& C_0^{1+u}(x,\omega)\left\{
  2(1-\alpha)(\alpha+u)+\frac{2\alpha (\alpha+u)}{1+\alpha-u} \right. \\ \nonumber
&+& \left. \left( \frac{2\alpha u(\alpha-u)}{1+\alpha -u} - 2(\alpha+u)\right)
 F\left(\frac{2\omega (1-x)}{1+\omega} \right) \right\}+ \{ x\leftrightarrow 1-x\}
 \ , \\ \nonumber
 R_3+\bar{R}_2 &=& C_0^{1+u}(x,\omega)\left\{
  2(1-\alpha)(\alpha+u)+\frac{2\alpha (\alpha+u)}{1+\alpha-u} \right. \\ \nonumber
&+& \left. \left( \frac{2\alpha u(\alpha-u)}{1+\alpha -u} - 2(\alpha+u)\right)
 F\left(\frac{-2\omega x}{1-\omega} \right) \right\}+ \{ x\leftrightarrow 1-x\}
 \ , \\ \nonumber
R_4+\bar{R}_4 &=& C_0^{1+u}(x,\omega) \frac{\alpha (\alpha + u) }
{C_0(x,\omega)\omega}
        \left\{ F\left( \frac{2\omega (1-x)}{1+\omega}  \right) -
             F\left( \frac{-2\omega x}{1-\omega} \right)   \right\} +
        \{ x \leftrightarrow 1-x \} \ . \nonumber
 \eeqn

From here, the subdivergences and the overall divergence have to be
subtracted. In \rcite{BB1} a technique to perform this in the $MS$
scheme was developed. We will now extend their method to arbitrary
schemes. First, one observes that a general
diagram with $n$ fermion blob insertions into the gluon
propagator can be expressed as \rref{allblobs}:

\be
D^{(n)} = (-1)^{n+1} C_F {\alpha _s \over 4 \pi}
{ (\alpha _s  \beta_0^f)^n \over (n+1)
\epsilon ^{n+1}  } E ( [n+1] \epsilon, \epsilon )
\rlabel{nbloobs}
\ee
where  $E ([n+1]\epsilon,\epsilon)$ is a
continuous function for $\epsilon =0$.
To subtract the subdivergences of the fermion bubbles, we also have
to consider diagrams where at least one of the fermion bubbles is
replaced by a counterterm:
\be
-C_F
{\alpha _s \over 4 \pi} \sum _{k=0} ^n { n \choose k }
{ (-\alpha _s \beta _0 ^f) ^{n-k} \over (n+1-k) \epsilon ^{n+1-k} }
E([n+1-k] \epsilon, \epsilon )
 \left( {\alpha _s \beta _0 ^f e^{\Omega \epsilon}
\over \epsilon} \right) ^k
\rlabel{nosubdivs}
\ee
In \rref{nosubdivs}, we have ($\ \hspace{-1mm} ^k _n$) diagrams
(permutations) containing $n-k$ fermion bubbles, \rref{nbloobs} and
$k$ counterterms. The subdivergences are cancelled by summing $k$
from 0 to $n$.
In the $MS$ scheme, only the poles in $\epsilon$ are subtracted, and
$\Omega _{MS} = 0$, but in other schemes also finite parts are
subtracted, and in what follows, we keep $\Omega \neq 0 $. The
structure of \rref{nosubdivs} suggests to define
\be
E([n+1] \epsilon, \epsilon ) = e^{(n+1)\Omega \epsilon}
E_\Omega ([n+1] \epsilon, \epsilon )
\ee
In terms of $ E_\Omega$, \rref{nosubdivs} reads
\be
- e^{(n+1)\Omega \epsilon} C_F
{\alpha _s\over 4 \pi} \sum _{k=0} ^n { n \choose k }
{ (-\alpha _s \beta _0 ^f) ^{n-k} \over (n+1-k) \epsilon ^{n+1-k} }
E _\Omega ([n+1-k] \epsilon, \epsilon )
 \left( {\alpha _s \beta _0 ^f
\over \epsilon} \right) ^k
\rlabel{nosubdivsBIS}
\ee
Now, $E ([n+1]\epsilon, \epsilon)$ is expanded in the following way:
\be
E_\Omega ([n+1]\epsilon, \epsilon) = \sum _{j=0} ^{\infty}
\tilde E _j (\epsilon) ([n+1] \epsilon) ^j, \qquad
\tilde E _0 (\epsilon) \equiv g(\epsilon)
= \sum _{i=0} ^{\infty} g _i
\epsilon ^i
\rlabel{series}
\ee
We now use \rref{series} to  expand
$E_\Omega ( [n+1-k] \epsilon, \epsilon )$. Keeping only terms that do
not vanish in the limit $\epsilon \rightarrow 0$, the contribution
of the $n$ fermion blob insertion, after subtraction of the
subdivergences, reads (see \rcite{BB1} for details):
\be
- e^{(n+1)\Omega \epsilon} C_F
{\alpha_s \over 4 \pi} { (\alfs \beta _0 ^f)^n  \over \epsilon ^{n+1} }
\left( { g (\epsilon) \over n+1}  + (-1)^n n!
\tilde E _{n+1} (\epsilon) \epsilon ^{n+1} \right)
\rlabel{noSUBDIVS}
\ee
Here, we have used
\be \sum _{k = 0} ^ n (-1)^{n-k}  ( n+1-k)^{J-1} = \left\{
\begin{array}{ll}
1/(n+1) & J = 0 \\
0 &0 < J < n + 1 \\
(-1)^n n! & J = n+1.
\end{array}
\right.
\ee
The second term in \rref{noSUBDIVS} is already UV finite and no
subtraction needs to be done, but the first one is UV divergent, and
the poles in $\epsilon$, together with some finite pieces have to
be subtracted. To subtract the finite pieces, a prescription has to
be fixed: we will cancel those finite pieces that come from the
expansion of the exponential. This is equivalent to taking
$\alpha _s e^{\Omega \epsilon}$, and not $\alpha _s$ as an expansion
parameter. The overall divergence can now easily be subtracted,
and we find the UV finite result
\be
-C_F
{\alpha_s \over 4 \pi}  (\alfs \beta _0 ^f)^n
\left( { g _{n+1} \over n+1}  + (-1)^n n!
\tilde E _{n+1} (0) \right)
\rlabel{UVFINITE}
\ee
In the first term
in \rref{UVFINITE}, $n$ can easily be summed from 0 to $\infty$:
\be
\sum _{n = 0} ^{\infty} -C_F
{\alpha_s \over 4 \pi } (\alpha_s \beta _0 ^f)^n
\left( { g  _{n+1} \over n+1} \right)
 = -{C_F \over 4 \pi \beta _0 ^f}
\int _0 ^{\alpha _s \beta _0 ^f} { g (\lambda) - g (0)
\over \lambda } d \lambda
\rlabel{convint}
\ee
To sum the second term, we use an integral representation for
$n!$:
\be - C_F
{\alpha _s \over 4 \pi }\int _0 ^\infty \sum _{n = 0} ^{\infty}
(- x \alpha _s \beta _0 ^f)^n e^{-x} \tilde E _{n+1} (0) dx =
{C_F\over 4 \pi \beta _0 ^f } \int _0 ^{\infty}
 e^{-u/(\alpha _s \beta _0^f)}
{E _\Omega (-u,0) - E _\Omega (0,0) \over u} d u
\rlabel{divint}
\ee
Applying this technique, we obtain:

\beqn
[D_i]_R &=&
  \frac{C_F}{ 4 \pi \beta_0^f}\left\{
\int\limits^\infty_0 e^{-u/\alfs \beta^f_0} \frac{du}u
 \left[\left(\frac{\mu^2e^{C}}{Q^2}\right)^{-u} \
 \frac{ 2\gamma_i(-u|x,\omega)}{(1+u)(2+u)}\
 -\gamma_i(0|x,\omega)\right] \right.\\ \nonumber
 &-& \left. \int\limits^{\alfs \beta^f_0}_0\
  \frac{d\lambda}{\lambda}
 [g_i(\lambda|x,\omega)- g_i(0|x,\omega)]\right\}\ ,  \nonumber
\eeqn
where
\beqn
g_i(\lambda|x,\omega) &=& \lim_{u\rightarrow 0}
E_i[u, \lambda] =
 \frac{\Gamma (4-2\lambda)}
{6(1-\lambda)\Gamma (1+\lambda)\Gamma^3(2-\lambda)}
R_i(\lambda,0|x,\omega) \ , \\ \nonumber
\gamma_i(-u|x,\omega) &=& \frac{(2+u)}{2(1-u)}
R_i(0,-u|x,\omega), \nonumber
\eeqn
and $C=5/3+\ln (4\pi)-\gamma_E + \Omega$
parametrizes the renormalizations scheme. In the $MS$ - scheme,
$ C=5/3+\ln (4\pi)-\gamma_E$ and in the $\overline{MS}$ - scheme,
$ C=5/3$.
Performing substitution $\beta_0^f \rightarrow  -b_0$
and rewriting the last integral in terms of
$G_i(1-\lambda|x,\omega)= g_i(\lambda|x,\omega) $ we obtain
our final result:
\beqn
[D_i]_R &=&
-\frac{C_F}{4 \pi b_0}\left\{
\int\limits^\infty_0 e^{-u/\alfs b_0} \frac{du}u
 \left[\left(\frac{\mu^2e^{C}}{Q^2}\right)^{u} \
 \frac{ 2\gamma_i(u|x,\omega)}{(1-u)(2-u)}\
 -\gamma_i(0|x,\omega)\right] \right.\\ \nonumber
 &+& \left. \int\limits^{1+\alfs b_0}_1\
  \frac{d\lambda}{1-\lambda}
 [G_i(\lambda|x,\omega)- G_i(0|x,\omega)]\right\}\ ,  \nonumber
\eeqn

\section*{Appendix B}

\setcounter{equation}{0}
\addtocounter{section}{1}
\setcounter{subsection}{0}

Here we present some technical remarks related to the solution of
the evolution
equation:
\begin{equation}
\left[ \mu^2\frac{\partial}{\partial \mu^2}
-b_0 \alfs ^2\frac{\partial}{\partial \alfs} \right]
\varphi(x,\mu^2)\ =\int^1_0 V_{\alpha}(x,y)\varphi(y,\mu^2)dy
\rlabel{eveq}
\end{equation}
Let us remind for convenience  some formulae for the eigenvalues
of the evolution kernel:
\begin{eqnarray}
\int^1_0
V_{\alpha}(x,y)\bar\varphi_{n}(y,\mu^2)&=&-\gamma_n(\alfs )
\bar\varphi_{n}(x,\mu^2)\ ,\nonumber \\
\bar\varphi_{n}(x,\mu^2)&=&
(x\bar x)^{\alpha}\varphi_{n}(x,\mu^2)A_{n}(\alfs )\ ,
\rlabel{eigen} \\
\varphi_{n}(x,\mu^2)&=& C^{1/2+\alpha}_{2n}(1-2x)\ ,\nonumber \\
 A_{n}(\alfs) &=&
\frac{\Gamma(1+2\alpha)}{\Gamma(\alpha)\Gamma(1+\alpha )}
\frac{(2n)!}{(2\alpha)_{2n} }
\frac{(1+2\alpha+4n)}{\alpha+n} \ ,\nonumber \\
\int^1_0\varphi_{n}\bar\varphi_{k}dx &=& \delta_{kn}\ ,
\rlabel{ort}
\end{eqnarray}
where $\alpha \equiv 1+b_0\alfs$.
It is natural to expand the  solution in a series of eigenfunctions
 \rref{eigen} of the evolution kernel:
\begin{equation}
\varphi(x,\mu^2)\ =\ (x \bar
x)^{1+\alfs b_0} \sum^\alpha_{n=0}
b_{n}(\mu^2)A_{n}(\alfs)C^{3/2+\alfs b_0}_{2n}(1-2x)\ .
\rlabel{decomp}
\end{equation}
Substituting \rref{decomp} in \rref{eveq} and using orthogonality
of the eigenfunctions
we obtain the evolution equation for the moments $b_{k}(\mu^2)$:
  \bee
D_{RG}b_{k}(\mu^2)+
  \sum^\infty_{n=0}b_{n}(\mu^2) \left\{
\varphi_{k}\otimes D_{RG}\bar \varphi_{n}\right\}
= -\gamma_{k}(\alpha)b_{k}(\mu^2).
\rlabel{momeq}
\een
where we have introduced the notation
\begin{eqnarray*}
D_{RG}\equiv
 \mu^2\frac{\partial}{\partial \mu^2}
-b_0 \alfs ^2\frac{\partial}{\partial \alfs}\, \\
\varphi_{k}\otimes \varphi_{n}\equiv
\int^1_0\varphi_{n}\varphi_{k}dx
\end{eqnarray*}
The second term in the lhs of \rref{momeq} arises due to the
dependence of $\bar \varphi_{n}$ on $\alfs(\mu^2)$.
We rewrite it in the following way:
\begin{eqnarray}
\sum^\infty_{n=0}b_{n}(\mu^2) \left\{
\varphi_{k}\otimes D_{RG}\bar \varphi_{n}\right\}=
\sum^\infty_{n=0}b_{n}(\mu^2)D_{RG} \left\{
\varphi_{k}\otimes \bar \varphi_{n}\right\}-
\sum^\infty_{n=0}b_{n}(\mu^2) \left\{
D_{RG}\varphi_{k}\right\}\otimes \bar \varphi_{n}
\rlabel{main}
\end{eqnarray}
The first term in the rhs of \rref{main} vanishes due to \rref{ort}.
Consider the second one.
\bee
D_{RG}\varphi_{k}=-b_0\alfs^2\frac{d}{d\alfs}C_{2k}^{3/2+b_0\alfs}(1-2x)
=-\sum^{k}_{\ell=0} C_{k \ell }(\alfs)C_{2\ell}^{3/2+b_0\alfs}(1-2x) \ .
\rlabel{sec}
\een
In order to calculate the coefficients $C_{k \ell}$,
it is convenient to
use the following formulae for the Gegenbauer polynomials \rcite{Chetyr}:
\begin{eqnarray*}
C^\nu_{2k}(t) &=& \sum^k_{\rho=0}\ \frac{(2t)^{2(k-\rho)}(-)^\rho\Gamma
(2k-\rho+\nu)}{(2[k-\rho])!\rho!\ \Gamma(\nu)}\ ; \\
\frac{(2t)^{2p}}{(2p)!} &=& \sum^p_{k'=0}\ C^\nu_{2p-2k'} (t)
\frac{\Gamma(\nu)(2p-2k'+\nu)}{\Gamma(k'+1)\Gamma(2p-k'+\nu+1)}\ .
\end{eqnarray*}
We have obtained:
\begin{eqnarray*}
C_{kk'}(\alpha) &=& (b_0\alfs)^2 \sum^k_{p=k'} \frac{(-)^{k-p}\
\Gamma(\frac32+b_0\alfs+k+p)}{(k-p)!(p-k')!\ \Gamma(\frac32+b_0\alfs+k'+p)}
\frac{(2k'+b_0\alfs+\frac32)}{(p+k'+b_0\alfs+\frac32)} \\
&\times& \left\{\psi\left(\frac32+b_0\alfs+k+p\right)-\psi\left(
b_0\alfs+\frac32\right)\right\}\ .
\end{eqnarray*}
In particular:
$$
  C_{kk}\ =\
(b_0\alfs)^2\left\{\psi\left(\frac32+b_0\alfs+2k\right)
   -\psi\left(b_0\alfs+\frac32\right)\right\},
   $$
Where $\psi(z)=\frac d{dz}\ln\Gamma(z)$.
Substituting \rref{sec} in \rref{main} and using \rref{ort}
we obtain:
\bee
 \left[ \mu^2\frac{\partial}{\partial \mu^2}-
 b_0\alfs^2\frac{\partial}{\partial \alfs}\right]
 b_{k}(\mu^2)=-b_{k}(\mu^2)\gamma_{k}(\alfs)-
 \sum^{k}_{\ell=0} C_{k \ell}(\alfs)b_{\ell}(\mu^2)\ .
\een

\end{document}